\documentclass[aps,floats,showpacs]{revtex4}
\usepackage{amsmath}
\usepackage{graphicx}

\begin{document}

\title{Electron lifetime in Luttinger liquids}
\author{Karyn Le Hur}
\affiliation{D\'epartement de Physique, Universit\' e de Sherbrooke, J1K 2R1 Sherbrooke, Qu\' ebec, Canada}

\date{\today}

\begin{abstract}
We investigate the decoherence of the electron wavepacket in purely ballistic one-dimensional systems described through the Luttinger liquid (LL). At a finite temperature $T$ and long times $t$, we show that the electron Green's function for a {\it fixed} wavevector close to one Fermi point decays as $\exp(-t/\tau_F)$ --- as opposed to the power-law behavior occurring at short times --- and the emerging electron lifetime obeys $\tau_F^{-1}\propto T$ for spinful as well as spinless electrons. For strong interactions,  $(T\tau_F)\ll 1$, reflecting that the electron is not a good Landau quasiparticle in LLs. We justify that fractionalization is the main source of electron decoherence for spinful as well as spinless electrons clarifying the peculiar electron mass renormalization close to the Fermi points. For spinless electrons and weak interactions, our intuition can be enriched through a diagrammatic approach or Fermi Golden rule and through a Johnson-Nyquist noise picture. We stress that the electron lifetime (and the fractional quasiparticles) can be revealed from Aharonov-Bohm experiments or momentum resolved tunneling. We aim to compare the results with those of spin-incoherent and chiral LLs.
\end{abstract}


\maketitle

%\email{Karyn. Le.Hur@USherbrooke.ca}

% text should be lower case, unless caps are necessary for meaning
%\keywords{Luttinger liquids, Dephasing of Aharonov-Bohm oscillations, Quantum noise}

\section{Introduction}

The understanding of electron decoherence in correlated systems is interesting in its own right because this reveals precious information about the fundamental physics of the electron-scattering mechanisms.
As a matter of fact, even though the interaction between electrons in conventional metals is strong, a single particle description still proves to be remarquably successful. This stems from the fact that elementary excitations are adiabatically connected to the electrons and can be described through {\it weakly interacting quasiparticles} (dressed electrons) resulting in the Fermi liquid. Its validity relies on the decay of electronic excitations being small as compared to the Fermi energy $E_F$. 

For Fermi liquid systems, the quasiparticle lifetime is the time $\tau_F$ for a dressed electron to redistribute its energy as a result of inelastic scattering events; a finite lifetime implies an uncertainty in energy of the quasiparticle. For times larger than $\tau_F$, quasiparticles
become so unstable that they lose any physical meaning. Consider the inelastic scattering rate of an excited (dressed) electron of energy $E$ with another electron involving an energy transfer $\omega$.  In $d>1$ dimensions Fermi Golden rule leads to the simple estimate \cite{Simons} (throughout the text $\hbar=1$ as well as $k_B=1$)
\begin{equation}
\frac{1}{\tau_F(E)} \sim \int_0^E \omega d\omega \int q^{d-1} dq \frac{|U_q|^2}{(q v_F)^2};
\end{equation}
energies are counted from the Fermi energy, $U_q$ denotes the screened Coulomb interaction, and
the factor $1/(q v_F)$ can be interpreted physically as the typical time an electron, with velocity $v_F$, spends in the interaction region. An assumption behind the Fermi liquid theory is the screening
of the Coulomb interaction resulting in $U_q\sim e^2/{(q^2+\eta^2)}$ where $\eta$ is the screening wavevector. Thus, in three dimensions, we check that $1/\tau_F$ is set by the phase volume
to be of order $E^2/E_F$ and $E_F$ is the Fermi energy whereas in two dimensions, one finds $1/\tau_F\propto E^2\ln(E/E_F)$ \cite{2D}. Since the Landau's quasiparticle picture is well justified when
$E\gg 1/\tau_F(E)$ this holds for two and three dimensions at sufficiently small energies or temperatures $(T)$. In a disordered conductor, the interaction time is set by the diffusion
time $\Re e[-i\omega+D{\bf q}^2]^{-1}$ and thus for any dimension $d$ this results in the important estimate $1/\tau_F\propto E^{d/2}$ \cite{Simons, Altshuler}.

On the other hand, electronic interference phenomena in solid-state systems and their destruction by
electron-electron interactions have been extensively studied the last decades. Examples include
weak-localization, Aharonov-Bohm oscillations in small rings, universal conductance fluctuations in
mesoscopic conductors, and the magnetoresistance of wires \cite{Imry}. In Fermi liquid systems, inelastic collisions are unambiguously the dominating source of phase-breaking (dephasing) at low temperatures. In one-dimensional (1D) and two-dimensional (2D) disordered conductors, the loss of phase coherence at low $T$ has been studied intensively, both theoretically  \cite{Imry,Altshuler,Altshuler2,Simons,Mirlin} and experimentally \cite{Marcus,Webb,Wind}.
In clean electron systems, the number of investigations are fewer. In two dimensions, experiments consistent with  the energy relaxation time $\tau_{\phi} \propto (T^2\ln T)^{-1}$ have been carried out in clean samples \cite{2Dexp}. Recent Aharonov-Bohm (AB) oscillations measured on {\it very clean} (ballistic) quasi 1D rings with only a {\it few propagating channels} support a dephasing time which varies as $\tau_{\phi}\propto T^{-1}$ and the dephasing of the electronic wavefunctions has been distinguished from thermal averaging effects \cite{Hansen}. The same results have been reproduced in a four-terminal geometry \cite{Kobayashi}. 

A theoretical endeavor to understand those experiments has been done by B\"{u}ttiker {\it et al.} \cite{Markus} by invoking the role of charge fluctuations between the 1D mesoscopic conductor (wire) and nearby gates. On the other hand, by assuming an explicit capacitive coupling (per length) $c$ between the mesoscopic wire and a side-gate, any deviation of the electron density from its mean value would then cost a certain energy $\sim e^2/c$; therefore, as pointed out in Ref. \cite{Blanter}, the gate induces an effective electron-electron interaction inside the wire. Of interest to us is thus to elucidate the notion of electron decoherence in a 1D very clean ballistic system with only one propagating channel by generically taking the electron-electron interaction into account, producing a Luttinger liquid ($T$ is much smaller than the electron bandwidth). We shall explain why the dephasing time of ballistic 1D conductors in Aharonov-Bohm experiments follows $\tau_{\phi}\propto T^{-1}$, considering the one channel case, that seems to be of experimental interest \cite{Hansen,Kobayashi}.

One very peculiar aspect of the LL is the breakdown of the Landau's quasiparticle picture and the
electron fractionalization mechanism implying that genuine charged excitations carry fractional quantum
numbers (for spinful but also for spin-polarized electrons); consult Ref. \cite{KV}. The LL state is, {\it e.g.}, characterized by  spin-charge separation realized by separate collective spin and charge excitations, each with its distinct propagation velocity. The tunneling density of states also exhibits a characteristic power-law suppression at low energies as a result of an ``orthogonality catastrophe'' at zero energy from the rearranging of the wave functions of the electrons to accommodate the new tunneling particle \cite{Glazman}. Experimental evidence for the LL state in 1D electron systems is by now irrefutable with measurements showing both the characteristic power-law suppression of the tunneling density of states and measurements of the spectral function providing direct evidence of spin-charge separation, including measures of the respective collective mode velocities \cite{Yacoby,Ishii,McEuen,Dekker}. Those experiments show the breakdown of the Landau's quasiparticle concept in LLs.  

In this paper, we are primarily concerned by the decoherence of the electron wavepacket in LLs. 
%We clarify that fractionalization is the main source of electron decoherence in (isolated) LLs for spinful %as well as spinless electrons: this results in a peculiar electron mass suppression close to the Fermi %level and this sheds some new light on the physical origin of the power-law suppression of the %tunneling density of states at low energy. 
More precisely, at a finite temperature $T$ and sufficiently long times $t$, we show that the electron Green's function for a {\it fixed} wavevector close to one Fermi point decays as $\exp(-t/\tau_F)$ for spin-polarized as well as spinful electrons and the electron lifetime obeys $\tau_F^{-1}\propto T$. On the other hand, for a given temperature, we will show that spin-charge separation strongly affects the interaction-dependent prefactor of the decoherence time. For strong interactions, this results in $(T\tau_F)\ll 1$ emphasizing the breakdown of the Landau Fermi liquid in 1D electron systems.   Expanding our previous Ref. \cite{KLH1}, we provide a physical justification of $\tau_F^{-1}\propto T$ in terms of the inherent electron fractionalization mechanism akin to Refs. \cite{KV,Safi}. For spinless electrons, a perturbative diagrammatic approach can also enrich our intuition \cite{Chubukov,Maslov,Gornyi}.  To second order in the electron-electron interaction the imaginary part of the self-energy effectively varies as $T^{-1}$. Moreover, this shows explicitly the peculiar electron mass suppression close to the Fermi points emphasizing that the electron is not a good quasiparticle in LLs. We stress that the electron lifetime can be detected through the dephasing of Aharonov-Bohm oscillations as observed in Refs. \cite{Hansen,Kobayashi} or via momentum-resolved tunneling \cite{wires,Fiete}. Keep in mind that the phase-breaking time of clean LLs, that controls the exponential suppression of electronic interferences as a function of the length of the interfering paths, is determined by the single-particle Green's function \cite{KLH1,KLH2,Gornyi}.

The paper is precisely organized as follows. In Sec. II, assuming spin-polarized electrons we develop the perturbative diagrammatic approach that provides a first justification of why the electron lifetime should be proportional to $1/T$ in LLs \cite{Maurice}. Again, this also clarifies that the electron acquires a singular mass at low energy traducing an underlying infra-red orthogonality catastrophe. In Sec. III,  by applying the Luttinger theory we build the electron lifetime from the exact Green's function at finite temperatures and flesh out the results by invoking the breakup of the electron wavepacket. We enrich our previous Ref. \cite{KLH1}. For weak interactions, we also formulate a Johnson-Nyquist noise description. In Sec. IV, we show explicitly the relevance of the electron lifetime on interference experiments --- expanding our Ref. \cite{KLH2} --- and on momentum resolved tunneling. We also comment on boundary and chirality effects. In Sec. V, we address the case of spinful electrons and in the low-density regime  \cite{M1} we discuss the crossover to the spin-incoherent LL \cite{FLB,FLB2}. Finally, appendices will be devoted to details of the calculations.

\section{A perturbative attempt for spinless electrons}

Firstly, we resort to a diagrammatic calculation akin to Refs. \cite{Chubukov,Maslov} to show that the perturbation theory entails some infra-red singularities but nevertheless for spinless fermions those (unphysical) singularities disappear when including vertex corrections or by exact cancellation between Hartree and exchange diagrams. This provides a first argument of why the electron lifetime in LLs is inversely proportional to $T$ even though the approach of Sec. II is restricted to the regime of vanishing interactions (we limit the calculations to second order in the Hubbard interaction).

\subsection{Power-counting of the self-energy}

\begin{figure}[htbp]
\begin{center}
\includegraphics[width=6.5cm]{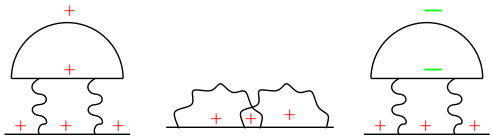}
\end{center}
\caption{(color online) Non-trivial self-energy diagrams to second order in $(Ua/v_F)$.}
\label{setup}
\end{figure}

One expects the following form for the imaginary part of the retarded self-energy ($W$ is a function of $|\omega|$ and $q$) 
\begin{equation}
\Im m\Sigma^R(\vec{k},E)\sim \int_0^{E} d\omega  \omega\int d^d q\  \Im m G^R(E-\omega,\vec{k}-\vec{q})W(|\omega|,q).
\end{equation}  
Here, we consider spin-polarized electrons. Interactions are supposed to be point-like, $d$ denotes the dimensionality of the system, $\Im m G^R$ refers to the imaginary part of the electron retarded Green's function, and the constraint on energy transfers $(0<\omega<E)$ is a manifestation of the Pauli principle limiting the number of accessible energy levels. 

The Fermi liquid assumption is as follows: when integrating over $q$ the result does not depend on $\omega$ resulting in
\begin{equation}
\Im m\Sigma^R(E)\sim C\int_0^E d\omega \omega \sim CE^2,
\end{equation}
where $C$ is the result of the $q$-integration that contains all the information about the interaction. Once we get the $E^2$-form for $\Im m\Sigma^R(E)$, the $E$-term in $\Re e\Sigma^R(E)$ immediately
follows from the Kramers-Kronig transformations, and we get a Fermi-liquid form of the self-energy regardless of a particular interaction and dimensionality. Thus a sufficient condition for the Fermi
liquid is the separability of the frequency and momentum integrations that can only happen if energy and momentum transfers are decoupled. Now, one can legitimately ask when and why this assumption is violated. A long-range interaction associated with small-angle scattering is known to destroy the Fermi
liquid \cite{Norman}.

In 1D ballistic systems, this assumption also breaks down since momentum conservation implies energy conservation. One way to perceive this is to focus on the
diagrams of Fig. 1. We introduce the lowerscripts $+$ and $-$ referring to right-moving and left-moving particles respectively. Second order in the Hubbard interaction $U$ leads to \cite{Maslov}
\begin{equation}
\hskip -0.5cm \Im m\Sigma_+^R(k,E) \sim (Ua)^2 \nu\sum_{j=\pm} \int_0^{E} d\omega  \int dq\  \Im m G_+^R(E-\omega,k-q)\Im m\Pi_{j}^R(\omega,q),
\end{equation}  
where $\nu$ is the density of states, $a$ the lattice spacing, the momenta will be measured from $+k_F$,  and 
$\Pi^R_{+}$ $(\Pi_-^R)$ is the retarded polarization bubble made of a right going (left going) electron. For a point-like interaction (within our notations) $\omega W(|\omega|,q)$ is proportional to the polarization bubble(s). For free  electrons, $\Im m G_+^R(E-\omega,k-q)=-\pi\delta(E-\omega-v_F(k-q))$, where $v_F$ is the Fermi velocity and the polarization bubbles take the explicit form  \cite{Giamarchi}
\begin{equation}
\Im m\Pi^R_{\pm}(q,\omega)= \pm\frac{q}{2} \delta(\omega\mp v_F q) = \frac{\omega}{2v_F} \delta(\omega\mp v_F q).
\end{equation}
At zero temperature, backward scattering of two electrons from different chiral (+ -) branches thus results in \cite{Maslov}
\begin{eqnarray}
\label{Imw}
\Im m\Sigma_{+}^R(k,E) \sim -(Ua)^2 \pi \nu \int_0^{E} d\omega  \int dq\  \delta(E-\omega+v_F q-v_F k)\frac{\omega}{2v_F} \delta(\omega+v_F q).
\end{eqnarray}
It is straightforward to get $-\left(Ua/v_F\right)^2 \pi E$ on the Fermi surface $(k\rightarrow 0)$. {\it Until the end of this subsection, we will continue to assume that $E>0$ and $k\rightarrow 0$}.
The Fermi Golden rule essentially leads to the same conclusion \cite{Gornyi}; $\omega$ is the energy gained by the other electron during the collision and the two $\delta$-functions reflect the momentum and energy conservations.  
%Momentum conservation automatically implies energy conservation resulting in $(k\rightarrow 0)$
%\begin{equation}
%\Im m\Sigma_+^R(E) \sim  -\left(\frac{Ua}{v_F}\right)^2\pi \int_0^E d\omega \omega \delta(E-2\omega). 
%\end{equation}
The linear behavior of $\Im m\Sigma_+^R(E,k\rightarrow 0)$ with $E$ irrefutably stems from the fact that momentum conservation also implies energy conservation.
It is certainly important to keep in mind the difference with the 1D disordered case \cite{Simons, Altshuler} where $q$ is homogeneous to $\sqrt{\omega/D}$ and thus  $\Im m\Sigma_+^R(E,k\rightarrow 0) \rightarrow  -(Ua)^2 \pi\nu\sqrt{E/D}$.

It is also interesting to note the $E/T$ scaling appearing in the results (to second order in $Ua/v_F$); for $E\ll T$ \cite{Chubukov}
\begin{eqnarray}
\Im m\Sigma_{+}^R(E-v_F k,T) \sim -(Ua)^2\nu\pi \int_0^{E} d\omega  \int dq\  \delta(E-\omega+v_F q-v_F k)\frac{\omega}{2v_F} \delta(\omega+v_F q)\coth\frac{E-v_F k}{4T},
\end{eqnarray}
(we have neglected $\tanh(E/4T)\rightarrow 0$) which then results in
\begin{equation}
\label{ImT}
\Im m\Sigma_{+}^R(k\rightarrow 0,E,T) \sim -\left(\frac{Ua}{v_F}\right)^2\pi \max\left(E,T\right).
\end{equation}
In the high-temperature regime, one thus expects the following form of the electron lifetime
\begin{equation}
\label{tauF}
\tau_F^{-1}[\alpha \ll 1] = -\Im m\Sigma_{+}^R(T) \sim \pi \alpha^2 T,
\end{equation}
where we have introduced the dimensionless parameter $\alpha=Ua/v_F=g_2/v_F$ where $g_2$ refers to backward scattering, {\it i.e.}, the scattering between two particles from different branches (+ and -) of the Fermi surface. 

More precisely, we extract the retarded Green's function
\begin{equation}
\label{free}
G_+^R(k,t) = -i\theta(t) e^{-ikv_Ft} e^{-t/\tau_F}. 
\end{equation}
Nevertheless, one might be dubatitive regarding this result: we have omitted the first two diagrams of Fig. 1 and we have neglected the real part of the self-energy. Below, we take those points precisely into account.

\subsection{On Forward Scattering}

The first two diagrams of Fig. 1 involve forward scattering $g_4$, {\it i.e.}, scattering between two particles from the same part of the Fermi surface (around $+k_F$); from the low-energy Luttinger theory, this forward scattering is well known to essentially produce a renormalization of the electron velocity $u>v_F$ (consult Appendix A) and thus this should not contribute to $\Im m\Sigma_+(E)$. The renormalization of the electron velocity can be seen by explicitly omitting the backward scattering $g_2$ (or the third diagram in Fig. 1) and by carefully estimating the vertex correction \cite{Giamarchi}
\begin{equation}
\Gamma_4=g_4-g_4\Pi_+^R(q,\omega)\Gamma_4,
\end{equation}
 resulting in:
\begin{equation}
\Gamma_4=\frac{g_4(\omega - v_F q)}{\omega-v_F q - \frac{g_4}{2\pi} q}=\frac{g_4(\omega-v_F q)}{\omega - u q}.
\end{equation}
This renormalized vertex now exhibits a pole at $\omega=uq=[v_F+g_4/(2\pi)]q$ instead of $\omega=v_F q$ reflecting the renormalization of the electron velocity. This renormalization of velocity also implies that $g_4$ does not contribute to $\Im m\Sigma_+(E)$. Indeed, by ignoring the backward scattering and by properly taking into account the vertex correction in Eq. (12) then the first diagram of Fig. 1 (on the mass shell $E=v_F k$) leads to the expected result:
\begin{eqnarray}
\Im m\Sigma_{+}^R(k,E) \sim -{g_4}^2 \pi \nu \int_0^{E} d\omega  \int dq\  \delta(E-\omega-v_F(k-q))\frac{\omega}{2v_F} \delta(\omega-u q) =0.
\end{eqnarray}
If one does not include the vertex correction then the first diagram would 
give $\Im m\Sigma_{+}^R(k,E) \propto E^2 \delta(E-v_F k)$; this yields a strong-delta-function-singularity on the mass shell $E=v_F k$ that is unphysical. However, for spinless fermions (by chance) the two first diagrams in Fig. 1 cancel each other exactly \cite{Gornyi,Maslov} and thus one might ignore vertex corrections and recovers the correct answer that for spinless fermions the forward scattering $g_4$ does not affect $\Im m\Sigma_+^R(k,E)$. 

Note that for spinful electrons those strong-delta-function-singularities in  $\Im m\Sigma_{+}^R(k,E)$ do not disappear and the forward scattering has to be treated in a 
non-perturbative manner to reproduce spin-charge separation (see Sec. V).

\subsection{Anomalous Mass}

From Kramers-Kronig relations, it is straightforward to obtain that the third diagram of Fig. 1 induces a real part to the self-energy, $\Re e\Sigma_+^R(E)$, that varies as $\alpha^2 E\ln(|E|/D)$  at zero temperature \cite{Maslov}. At a general level, we will denote $D$ the ultraviolet cutoff of the theory; for the present case, one can safely replace $D\sim v_F/a$ by the Fermi energy $E_F={k_F}^2/(2m)$ of the electron liquid. In the quantum realm, the corresponding electron spectral weight satisfies
\begin{equation}
\label{Zek}
{\cal Z}(k,E)\approx \left(\frac{E-v_F k}{D}\right)^{\alpha^2}.
\end{equation}
The electron becomes dressed by a large number of electron-hole pairs, that will lead to the electron fractionalization scheme described below, and this in turns drives the electron spectral weight to zero on the mass shell $(E=v_F k)$. Keep in mind the anomalous mass enhancement  becoming singular at low $E\ll \Delta=D\exp(-1/\alpha^2)$: $m/m^*={\cal Z}(k\rightarrow 0,E)\approx (E/D)^{\alpha^2}$.
%\begin{equation}
%\frac{m}{m^*(E)} \approx 1+\frac{\partial\Re e\Sigma_+^R(E)}{\partial E} \approx
%1+\alpha^2 \ln(E/D)\approx \left(\frac{E}{D}\right)^{\alpha^2}.
%\end{equation}
This traduces the non-applicability of the Landau Fermi liquid in low dimensions.  We infer
\begin{equation}
G_+^R(k,E) \approx  \left[E-v_F k\right]^{\alpha^2 -1} D^{-\alpha^2}.
\end{equation}
The latter reveals an inherent branch cut at $E=v_F k$ instead of a quasiparticle pole! At zero temperature, it is relevant to note that $\Im m\Sigma_+(E)$ now becomes negligible in front of the anomalous mass effect. This results in
\begin{eqnarray}
\label{pt}
G_+^R(p,t) = -i\theta(t)\left(\frac{a}{t v_F}\right)^{\alpha^2} e^{-i(p-k_F)v_Ft}.
\end{eqnarray}
Increasing the temperature $T\rightarrow E$, we rather estimate
\begin{equation}
G_+^R(k,E)\approx \frac{(T/D)^{\alpha^2}}{E-v_F k +i\pi\alpha^2 T},
\end{equation}
thus leading to:
\begin{eqnarray}
\label{ppt}
G_+^R(p,t) &=& -i \theta(t)\left(\frac{T}{D}\right)^{\alpha^2} e^{-i(p-k_F)v_Ft}\exp(-\pi\alpha^2T t).
\end{eqnarray}
{\it The two distinct behaviors of $G_+^R$ with time will be explicitly reproduced from the exact electron Green's function}. Moreover, from the correspondence $t\leftrightarrow 1/E$, the crossover between Eqs. (\ref{pt}) and (\ref{ppt}) should arise when $t T \approx 1$.

\section{Electron lifetime from the exact Green's function}

Nevertheless, the perturbative calculation entails some non-perturbative effects
close to $E=v_F k$ (see {\it e.g.} Eq.  (15)) and a non-perturbative treatment is suitable to make more quantitative predictions. We argue that the physics is more readily revealed by bosonization (see Appendix A); for more details on the method, consult, {\it e.g.}, Ref. \cite{Giamarchi}.

\subsection{Electron Green's function}

In Appendix A, we also provide a pedestrian derivation (based on the Luttinger theory) of the time-ordered electron Green's function for a wire with length $L\rightarrow +\infty$ and finite temperature $(\beta=1/T)$. The final result is
\begin{eqnarray}
\label{Gfinite}
G_+(x,\tau>0) &=& -e^{i k_F x}\langle T_{\tau} \Psi_+(x,\tau)\Psi^{\dagger}_+(0,0)\rangle \\ \nonumber
&=& - e^{ik_F x}\frac{a^{2\gamma}}{2\pi}\left(\frac{\pi}{u\beta}\right)^{2\gamma+1} \left(\sin\left[\frac{\pi}{\beta}\left(\tau-i\frac{x}{u}\right)\right]\right)^{-\gamma-1}\left(\sin\left[\frac{\pi}{\beta}\left(\tau+i\frac{x}{u}\right)\right]\right)^{-\gamma},
\end{eqnarray}
where $\tau=it$ is the Matsubara time and for convenience we introduce the useful parameter
\begin{equation}
\gamma=-1/2+(g+g^{-1})/4>0.
\end{equation} 
The Green's function of Eq. (19) is in complete accordance with that of Ref. \cite{Gornyi}. It is appropriate to note that for reasonable interactions $\gamma$ is always much smaller than one ($\gamma=0$ for free electrons and $\gamma=1/8$ for infinite on-site interactions producing $g=1/2$). The main contribution of $G_+(p,t)=\int dx\ e^{-ipx}G_+(x,t)$ will stem from the (branch cut) region around $x=ut-i0^+>0$ where $u$ denotes the renormalization of the (plasmon) velocity. 

For quite long times $2\pi t/\beta\gg 1$, by exploiting Fig. 2, we thus converge to:
\begin{eqnarray}
\label{p}
G_+(p,t) \approx -i\int_{-\infty}^{+\infty} dx e^{-i(p-k_F)x}
\left[(-i)(x-ut+i0^+)\right]^{-1-\gamma}a^{\gamma}\left(\frac{a\pi}{u\beta}\right)^{\gamma}\exp\left[-\frac{\pi}{\beta}2\gamma t\right].
\end{eqnarray}
Thus, this results in (assuming that $p>k_F$ for the electron)
\begin{eqnarray}
\label{tauF}
G_+^R(p,t) \approx -i\theta(t)\left[a(p-k_F)\right]^{\gamma} \left(\frac{Ta}{u}\right)^{\gamma} e^{-i(p-k_F)ut} e^{-2\pi \gamma t/\beta}.
\end{eqnarray}
Unambiguously, we extract the electron lifetime
\begin{equation}
\tau_F^{-1} = 2\pi \gamma T = \pi T\left[\frac{(g+g^{-1})}{2}-1\right].
\end{equation}

\begin{figure}[htbp]
\begin{center}
\includegraphics[width=6.5cm,height=4.8cm]{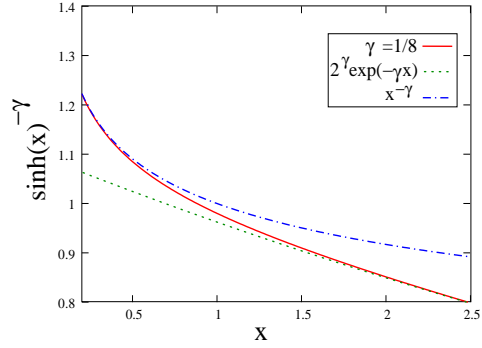}
\end{center}
\caption{(color online)  Behavior of $\sinh^{-\gamma}(x)$ (in the text $x=2\pi t\beta$) with the
two asymptotes $x^{-\gamma}$ at short times and $2^{\gamma}\exp(-\gamma x)$ at long times. For finite $\gamma$ and large times such that $t>\tau_F \gamma$, the Green function $G_+(p,t)$ falls off exponentially with time.}
\label{setup}
\end{figure}

For $Ua\ll v_F$, we get $g\sim 1-Ua/(2\pi v_F)$, allowing us to confirm the perturbative form of the electron lifetime, $\tau_F^{-1}[\alpha \ll 1] \propto \pi T(Ua/v_F)^2$.  The calculation of Sec. II, however, does not reproduce the prefactor $\left[a(p-k_F)\right]^{\gamma}$; this is not so surprising because the latter stems from a non-perturbative effect, {\it i.e.}, the fractionalization of the electron wavepacket as discussed below Eq. (35). More precisely, the perturbative analysis is equivalent to approximate $(x-ut)^{-1-\gamma}\sim (x-ut)^{-1}$ in Eq. (19) but this is not too bad since $\left[a(p-k_F)\right]^{\gamma}\rightarrow 1$ when $\gamma\rightarrow 0$. In the crystal limit $(g\ll 1)$, $(T\tau_F)\ll 1$, emphasizing that in LLs the electron is not a good Landau quasiparticle at low energy.

In the short-time limit $2\pi tT\leq 1$, the electron Green's function gets modified as:
\begin{equation}
G_+^R(p,t) \approx  -i\theta(t)e^{-i(p-k_F)ut} 
\left[a(p-k_F)\right]^{\gamma}\left(\frac{D t}{a}\right)^{-\gamma}.
\end{equation}
{\it In agreement with the perturbative approach of Sec. II C, the electron lifetime cuts off the power-law decay with time of the electron Green's function}. In the quantum limit $T\rightarrow 0$, this allows us to recover the well-known result $(E>0)$:
\begin{equation}
\label{Gpe}
G_+^R(p,E) = \int dt e^{iEt} G_+^R(p,t) \approx  \left[a(p-k_F)\right]^{\gamma} \left[E-u(p-k_F)\right]^{\gamma -1} D^{-\gamma}.
\end{equation}
For reasonable interactions, {\it i.e.}, assuming that we are always in the limit $E\ll \Delta=D\exp(-1/\gamma)$, the branch cut occurring at $E=u(p-k_F)$ now becomes very prominent. We infer the
following spectral function
\begin{equation}
{\cal A}_+^R(p,E\rightarrow 0)=-\Im m G_+^R(p,E\rightarrow 0)\rightarrow (E/D)^{2\gamma}\delta(E-u(p-k_F)).
\end{equation}
In passing, we also check that:
\begin{equation}
G_+^R(x\rightarrow 0,t) = \int dp G_+^R(p,t) \propto \left(\frac{a}{D t}\right)^{2\gamma+1}.
\end{equation}
We like to mention that the bulk exponent $2\gamma+1=(g+g^{-1})/2$ must be explicitly distinguished from the edge exponent $g^{-1}$ \cite{Glazman}. In the finite temperature realm and $E\rightarrow 0$ (or times $2\pi Tt\geq 1$), we recover a Lorentzian peak:
\begin{equation}
\label{high}
G_+^R(p,E\rightarrow 0,T) \approx \frac{T^{2\gamma}D^{-2\gamma}}{E-u(p-k_F)+i2\pi\gamma T}.
\end{equation}
In the crossover realm $E\approx T$, one can substitute $T^{2\gamma}\rightarrow T^{\gamma}[\max(E,T)]^{\gamma}$ in Eq. (28).

At this point, we like to specify that similar results can be derived for left-moving electrons and in
particular
\begin{eqnarray}
G_-(x,\tau>0) &=& -e^{-i k_F x}\langle T_{\tau} \Psi_-(x,\tau)\Psi^{\dagger}_-(0,0)\rangle \\ \nonumber
&=& - e^{-ik_F x}\frac{a^{2\gamma}}{2\pi}\left(\frac{\pi}{u\beta}\right)^{2\gamma+1} \left(\sin\left[\frac{\pi}{\beta}\left(\tau+i\frac{x}{u}\right)\right]\right)^{-\gamma-1}\left(\sin\left[\frac{\pi}{\beta}\left(\tau-i\frac{x}{u}\right)\right]\right)^{-\gamma}.
\end{eqnarray}

\subsection{Fractionalization}

\begin{figure}[htbp]
\begin{center}
\includegraphics[width=10.0cm]{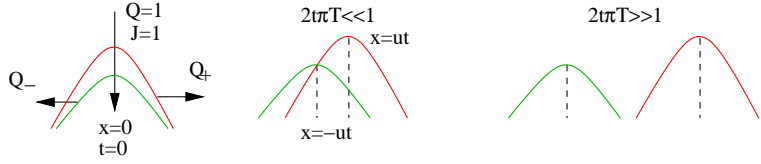}
\end{center}
\caption{(color online) For $2\pi t\ll \beta$, although the overlap between the fractional wavepackets is finite, the inherent fractionalization process clarifies the orthogonality catastrophe factor $t^{-\gamma}$ in ${\cal G}_+(x\rightarrow ut,t)$ whereas at long times the electron fractionalization leads to ${\cal G}_+(x\rightarrow ut,t)\propto \exp(-t/\tau_F)$ with $\tau_F^{-1}=2\pi\gamma/\beta$. For very weak interactions, we obtain $\tau_F^{-1}\propto (Ua/\pi v_F)^2 T$.
}
\end{figure}

Below, we provide a relatively simple interpretation of the results based on electron fractionalization.
More precisely, let us inject an electron with momentum $+k_F$ exactly at the coordinate $x=0$ and time $t=0$: exploiting Refs. \cite{KV,Safi} and Appendix B, we infer that there will be fractionalization of the electron wavepacket into a charge $Q_+=(1+g)/2$ state going to the right and a charge $Q_-=(1-g)/2$ going to the left. Those fractional charges must be distinguished from Laughlin quasiparticles with charges $g$ that are rather produced by the backscattering from impurities \cite{Glazman,Trauzel}.

It is thus convenient to identify (we find it appropriate to keep the same notations as in our Ref. \cite{KLH1})  
\begin{equation}
\Psi_+^{\dagger}(0,0)=\frac{1}{\sqrt{2\pi a}}\exp \left[-i\sqrt{\pi}(-\phi+\theta)(0,0)\right]
= \frac{1}{\sqrt{2\pi a}}{\cal L}_+^{\frac{1+g}{2}}(0,0){\cal L}_-^{\frac{1-g}{2}}(0,0).
\end{equation}
The creation operators ${\cal L}_{\pm}^{Q_{\pm}}$ (with the lowerscript $\pm$ referring to the
direction of propagation and $Q_{\pm}$ referring to the associated fractional charge) are precisely given
in Appendix B. Moreover, by exploiting Eq. (A29), we predict
\begin{eqnarray}
<({\cal L}_+^{\frac{1+g}{2}})^{\dagger}(x,\tau)
{\cal L}_+^{\frac{1+g}{2}}(0,0)> &=& \exp\left(\pi\frac{(1+g)^2}{4}<\theta_+(x,\tau)\theta_+(0,0)-
\theta_+(x,\tau)^2>\right)\\ \nonumber
&=&  \left[\frac{a\pi}{u\beta}\right]^{\frac{(1+g)^2}{4g}}
\left[\sin\left(\frac{\pi}{\beta}[\tau-i\frac{x}{u}]\right)\right]^{-\frac{(1+g)^2}{4g}},
\end{eqnarray}
as well as
\begin{equation}
<({\cal L}_-^{\frac{1-g}{2}})^{\dagger}(x,\tau)
{\cal L}_-^{\frac{1-g}{2}}(0,0)>\ =  
\left[\frac{a\pi}{u\beta}\right]^{\frac{(1-g)^2}{4g}}\left[\sin\left(\frac{\pi}{\beta}[\tau+i\frac{x}{u}]\right)\right]^{-\frac{(1-g)^2}{4g}}.
\end{equation}
This clarifies the form of the electron
Green's function (note that $(Q_-)^2/g=(1-g)^2/(4g)=\gamma$ and $(Q_+)^2/g=1+\gamma$): 
\begin{equation}
<\Psi_+(x,\tau)\Psi_+^{\dagger}(0,0)>\ =\ \frac{1}{2\pi a} 
<({\cal L}_+^{\frac{1+g}{2}})^{\dagger}(x,\tau){\cal L}_+^{\frac{1+g}{2}}(0,0)>
\times<({\cal L}_-^{\frac{1-g}{2}})^{\dagger}(x,\tau){\cal L}_-^{\frac{1-g}{2}}(0,0)>.
\end{equation}
{\it We emphasize that this identification is in fact 
essential to correctly interpret the electron Green's function in LLs}. 

More precisely, the right-moving fractional charge $Q_+=(1+g)/2$ propagates at the plasmon sound
velocity $u$; one can check that its Green's function indeed possesses a resonance
for $\tau=ix/u$ $(t=x/u)$. Thus, in a time $t>0$,
the charge $Q_+$ will be located at a certain position $x\approx ut$. On the other hand, 
the probability amplitude that the
counter-propagating charge $Q_-= (1-g)/2$ reaches the same position at the time $t$ such
that $2t\pi/\beta\gg 1$ obeys
\begin{equation}
<({\cal L}_-^{\frac{1-g}{2}})^{\dagger}(x,t)
{\cal L}_-^{\frac{1-g}{2}}(0,0)>|_{x\rightarrow ut>0}
\propto \left[
\sinh\left(\frac{\pi}{\beta}2t\right)\right]^{-\gamma}
\approx\exp(-t/\tau_F),
\end{equation}
where $\tau_F^{-1}=\pi2\gamma/\beta$ is precisely the electron lifetime defined in Sec. III A. In particular, this gives a clear justification to the exponential decay of the electron Green's function 
at long times $2\pi t/\beta\gg 1$ (consult Eq. ({\ref{tauF}) and Fig. 3), {\it i.e.},
\begin{equation}
{\cal G}_+(x\rightarrow ut,t)=(x-ut)^{1+\gamma}G_+(x\rightarrow ut,t)
\rightarrow -i\theta(t)e^{ik_Fut}\exp\left(-2\gamma\pi t/\beta\right). 
\end{equation}
Keep in mind that the charge propagating to the right is fractional; more precisely, $\gamma +1 = (Q_+)^2/g$, and this
 is at the origin of the ``orthogonality catastrophe'' factor $[a(p-k_F)]^{\gamma}$ in Eq. (\ref{tauF}) with  $\gamma = (Q_-)^2/g=(Q_+)^2/g -1$.

\subsection{Johnson-Nyquist noise picture for weak interactions}

It has been emphasized since more than a decade that a fluctuating potential $V(t)$ with Johnson-Nyquist type correlations can mimic the electronic interactions in a one-channel mesoscopic conductor \cite{Nazarov,Girvin}. Such a correspondence in one dimension has been well established in a {\it point-like} model, {\it e.g.}, for a tunnel barrier \cite{IH} or a quantum dot \cite{KM} and we propose to discuss it in the context of the electron Green's function $G_+^R(p,t)$ in a clean and infinite wire.

In the classical regime $T\gg E$, the Johnson-Nyquist noise obeys  $(\langle V(t)\rangle=0)$:
\begin{equation}
\langle V(t) V(0) \rangle = 2\pi r\beta^{-1}\delta(t),
\end{equation}
where $r$ is a fictitious dimensionless parameter related to the strength of the interactions.

Exploiting Appendix A, we include an extra dissipative term in the Lagrangian of the form:
\begin{eqnarray}
\delta {\cal L} = \int dx\ \frac{1}{\sqrt{\pi}}\partial_x\phi(x) V(t) = \frac{1}{v_F\sqrt{\pi}}\int dx\ \partial_t \theta V(t).
\end{eqnarray}
We have used that $\partial_t\theta=i[{\cal H},\theta]=v_F\partial_x\phi$. Owing to the fact that a free electron gas is embodied by the Lagrangian
\begin{equation}
{\cal L} = \frac{1}{2}\int dx\ \frac{1}{v_F}\left(\partial_t \theta\right)^2-v_F\left(\partial_x\theta\right)^2,
\end{equation}
$\delta {\cal L}$ can be absorbed into ${\cal L}$ by modifying $\sqrt{\pi}\dot{\theta}\rightarrow \sqrt{\pi}\dot{\theta}+ V(t)$. Now, from Eq. (A1), we infer that this will modify:
\begin{eqnarray}
\label{trente}
G_+^R(p,t) \approx -i\theta(t) e^{-i(p-k_F) v_F t} e^{i{\cal K}(t)},
\end{eqnarray}
where ${\cal K}(t)=\int_0^t  \ dt' V(t')$. Moreover, modeling the Johnson-Nyquist noise by  a set of harmonic oscillators \cite{Markus2, MB}, {\it i.e.}, exploiting the identity $\langle e^{i{\cal K}(t)} \rangle = e^{-\langle {\cal K}(t)^2 \rangle /2}$, and averaging $G_+(p,t)$ over the  set of harmonic oscillators we infer
\begin{eqnarray}
\label{trente2}
\langle G_+^R(p,t) \rangle \approx -i\theta(t) e^{-i(p-k_F)v_F t} e^{-\pi r t/\beta}.
\end{eqnarray}
We have exploited $\langle {{\cal K}(t)}^2\rangle=\int_0^{t} dt' \int_0^{t} dt'' \langle V(t') V(t'') \rangle =2\pi r\beta^{-1} t$. Having in mind the results of Sec. III, this strongly suggests the identification $r=2\gamma$; this  differs from the equality  $r=(g^{-1}-1)$ valid at the edge of a wire \cite{IH}. 

However, at this point, let us stress that Eq. (\ref{trente2}) is only valid for very weak interactions where one can safely approximate $u\approx v_F$ and $\Delta=D\exp(-1/\gamma)\rightarrow 0$; Eq. (\ref{trente2}) becomes completely equivalent to Eq. (22). Assuming weak interactions, the analogy with the Johnson-Nyquist noise can be made more explicit by deriving explicitly the phase uncertainty
%\begin{equation}
%\langle {{\cal K}(t)}^2\rangle \propto -(Ua/v_F)^2\beta^{-1}\int_0^{t} dt' \int_0^{t} dt'' \int %\frac{d\omega}{2\pi} \int \frac{dq}{2\pi} \frac{1}{\omega}\Im m \Pi^R_-(q,\omega) e^{iq[x(t')-x(t'')]}e^{-%i\omega (t'-t'')},
%\end{equation}
%with $\Im m\Pi_-^R(q,\omega) \approx \omega\delta(\omega+v_F q)$ and for the ballistic case $x(t')-%x(t'')=v_F(t'-t'')$ leading to 
$\langle {{\cal K}(t)}^2\rangle \approx (Ua/v_F)^2\beta^{-1} t$ from the diagrammatic approach of Sec. I. For $T\rightarrow 0$, one gets $\langle {{\cal K}(t)}^2\rangle=-2r\ln(tD)$ \cite{KLH2} allowing us to recover the local Green's function $G_+^R(x\rightarrow 0,t)=\int dp G_+^R(p,t) \propto t^{-2\gamma-1}$.

%The extra phase ${\cal K}(t)$ in Eq. ({\ref{trente}}) will have important consequences for  Aharanov-%Bohm type experiments resulting in the dephasing of mesoscopic interferences. 
It is relevant to observe that the Johnson-Nyquist noise formulation provides a precious link between electron-electron interactions and dephasing processes \cite{KLH2}. The dephasing time appearing in mesoscopic interferences will be $\tau_{\phi}^{-1}\approx \tau_F^{-1}=2rT$. For weak interactions, this results in $\tau_{\phi}^{-1}\approx (1-g^2)^2 T$ in accordance with Ref. \cite{Markus}.

\section{How to probe the electron lifetime}

Now, we address the important question if whether the electron lifetime can be observed experimentally.  At this step, it is important to recall that the current through a (single) quantum wire does not reveal the electron lifetime and by including the reservoir leads results in a quantized conductance (per transverse mode) $\sim 2e^2/h=1/\pi$ when assuming a perfectly clean wire \cite{Maslov}. To recover such a result, one can make use of the fractionalization formalism in Appendix B leading to a Fabry-Perrot resonator of fractional wavepackets \cite{Safi}. Below, we propose simple transport experiments, such as momentum resolved tunneling or Aharanov-Bohm type oscillations from two weakly-coupled wires, that might probe the exponential decay of the single-electron Green's function with
distance (time) explicitly.

\subsection{Momentum resolved tunneling density of states}

\begin{figure}[htbp]
\begin{center}
\includegraphics[width=7.5cm,height=2cm]{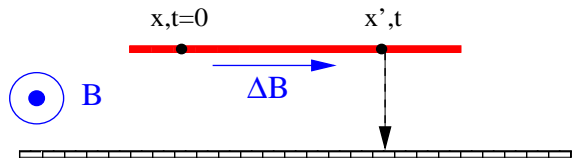}
\end{center}
\caption{(color online) Schematic geometry of electrons tunneling between the wire and the source.
A magnetic field $B$ is applied perpendicular to the plane of the wire to allow a field-dependent momentum boost  $p_B=B d$ where $d$ is the distance between the wire and the source \cite{wires,Fiete}. Electrons can be made ``spinless'' through another magnetic field $\Delta B$ applied along the  wire.}
\label{setup}
\end{figure}

Here we discuss the effect of the results above on the momentum resolved tunneling density of states \cite{Yacoby,Fiete,wires}
\begin{equation}
N_+(p)=\frac{1}{L}\int dx \int dx' e^{-ipx'}e^{ipx} \Psi_N(x') \Psi_N^*(x),
\end{equation}
with the ``quasi-wavefunction'' $\Psi_N(x)=e^{ik_F x}\langle \Psi_{N-1}|\Psi_+(x)|\Psi_{N}\rangle$; in the absence of interactions, $\Psi_N(x)$ would be the wavefunction of the Nth electron, {\it i.e.}, a plane-wave since below we assume an infinite wire. For convenience, we limit the discussion to momenta close to $+k_F$ but similar features are found at $-k_F$. Applying the Green function formalism above, the momentum resolved tunneling density of states $N_+(p,V,T)$
can be evaluated through the Green's function $G_+(x'-x,V,T)$ with $V>0$ being the applied voltage between the wire and the source \cite{Fiete}. For free electrons, one gets $N_+(p,V) = \delta(V-u(p-k_F))$ \cite{Fiete} whereas the primary effect of the interaction when $T\rightarrow 0$ is to produce a power-law suppression $N_+(p\rightarrow k_F)\approx \left(V/D\right)^{2\gamma}$ due to
the breakup of the electron wavepacket; this can be explicitly obtained from Eq. (26) by taking the limit $E=V\rightarrow 0$ leading to $G_+^R(p,V\rightarrow 0)\sim -i (V/D)^{2\gamma}\delta(V-u(p-k_F))$. Note, $2\gamma=-1+(g+g^{-1})/2=2(Q_-)^2/g$ is the exponent for bulk tunneling (as opposed to $g^{-1}-1$ at the edges) \cite{Glazman}. 

Let us now enhance the temperature up to $T\geq V$. Exploiting the electron Green's function of Eq. (28) results in
\begin{equation}
\label{density}
G_+(x'-x, V\leq T) \approx \frac{-i}{u}\theta(x'-x)e^{i k_F(x'-x)} e^{i \frac{V}{u}(x'-x)}
[T/D]^{2\gamma} \exp\left[-\frac{(x'-x)}{u\tau_F}\right],
\end{equation}
and therefore explicitly in:
\begin{equation}
N_+(p,V\rightarrow 0,T)=\Re e\int d(x'-x) i G_+(x'-x,V\rightarrow 0,T) e^{-ip(x'-x)} =\frac{1}{D}
\frac{\tau_F^{-1} T^{2\gamma} D^{-1-2\gamma}}{\left[a(p-k_F)\right]^2 + \left[\tau_F D\right]^{-2}}.
\end{equation}
As shown in Fig. 4, the $\delta$-function at $p=+k_F$ or $k=(p-k_F)=0$ is broadened by an amount 
$\delta k = \frac{1}{u}\tau_F^{-1} = 2\pi\gamma T/u$ converging to $\pi T/(2v_F)$ for sufficiently strong interactions. {\it For spin-polarized electrons, it is relevant to note that owing to $T\ll D$, the peak at $k_F$ is narrow, {\it i.e.}, the broadening of the peak always remains much smaller than $k_F$}.

\begin{figure}[htbp]
\begin{center}
\includegraphics[width=8cm,height=5.2cm]{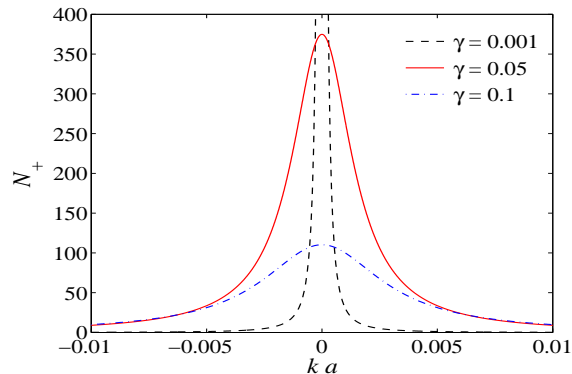}
\end{center}
\caption{(color online) Broadening of the momentum resolved tunneling density of states by electron-electron interactions for an infinite and spin-polarized wire; $D=1$ and $T=0.05$. The width of the peak gives an access to $\gamma/u=(Q_-)^2/(gu)=(Q_-)^2/v_F$.}
\label{setup}
\end{figure}

Remember that the momentum resolved density of states allows in principle to explicitly reveal the electron lifetime;  this irrefutably traduces that the electron Green's function in position space falls off exponentially with distance.

\subsection{Electronic Interferences}

\begin{figure}[htbp]
\begin{center}
\includegraphics[width=6.5cm,height=4cm]{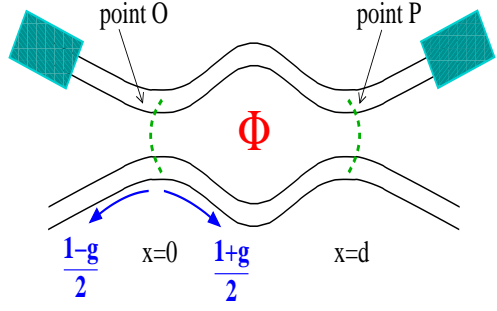}
\end{center}
\caption{(color online) The setup for electron wave interferences. The electrons can tunnel from one wire to the other at $x=0$ and $x=d$. The lengths of the wires are much larger than $d$ such that fractionalization can not be hindered by boundary effects.}
\end{figure}

Now, we focus on the setup of Fig. 5 composed of two weakly tunnel-coupled wires. We like to mention that the form of the weak tunnel coupling is explicitly shown in Appendix C.
For simplicity, we consider the situation of symmetric wires but the results can be easily generalized to wires with different interaction strengths or electron densities. The conductance of the upper wire now contains an additional interference term that stems from the ability of an electron in the upper (1) wire to escape into the lower (2) wire at the site $x=0$ and to return into the upper wire at the other tunneling point $x=d$; $d$ being the distance between the two tunneling positions (denoted 0 and P in Fig. 5). This contribution to the current in the upper wire depends on the enclosed magnetic flux $\Phi$. The {\it main part} of the flux-dependent part of the current has been computed in our Ref. \cite{KLH2} and can be written as \cite{noteI}
\begin{equation}
I_{\Phi} \approx -2\frac{e^2}{h}|\Gamma_0 \Gamma_d| u^2 \Im m\left[e^{2i\pi\Phi/\Phi_o} X_{d0}^R(\omega)+h.c.\right]_{\omega=V},
\end{equation}
where $X_{d0}^R(\omega)$ is the Fourier transform  of $X_{d0}^R(t)=-i\theta(t)\langle[B_+(d,t),B_+^{\dagger}(0,0)] \rangle$ and $B_+(x,t) = \Psi_{+2}(x,t)\Psi_{+1}^{\dagger}(x,t)$ embodies an electron tunneling operator acting at point $x$. 
{\it In the relevant regime $Td/u\approx 1$, the tunneling operator involving the backscattering of an electron is much less important as shown in Sec. IV. C and thus can be neglected.}

We denote $V$ the bias voltage applied through the upper wire and we choose the gauge such that the electrochemical potential of right-movers is $V$ whereas
the electrochemical potential of left-movers is zero;  another choice of gauge would lead to the same physical result. Moreover, $\Gamma_0$ and $\Gamma_d$ denote the
(dimensionless) tunneling amplitudes at the two tunneling positions and $\Phi_o$ is the flux quantum. We consider the realm $D\gg T\gg V$ ($V\rightarrow 0$). It is relevant to note that there is another energy scale in the problem, $u/d$, that stands for the distance between the tunneling points. 

Using the electron Green's function for the important realm $T\geq u/d$ leads to (details are given in Appendix C):
\begin{equation}
\label{AB}
I_{\Phi} = \frac{e^2 V}{h} |\Gamma_0 \Gamma_d | \cos(2\pi\Phi/\Phi_o)\left(\frac{T}{D}\right)^{g+g^{-1}-2} \exp\left[-\frac{2\pi \gamma T(2d)}{u}\right].
\end{equation}
When the distance $d$ is large enough $(Td/u\gg 1)$, the electron fractionalization at $x=0$ (Fig. 5) now results in:
\begin{equation}
G_{\Phi}=dI_{\Phi}/dV=\frac{e^2}{h} |\Gamma_0 \Gamma_d| \cos(2\pi\Phi/\Phi_o)\left(\frac{T}{D}\right)^{4\gamma} \exp[-2d/(u\tau_F)].
\end{equation}
%Note that the Luttinger theory is still applicable for those temperatures when assuming $d\gg a$ %implying $T\ll D$.
The mesoscopic interference visibility becomes explicitly suppressed as $(T/D)^{4\gamma}\exp(-2d\tau_{\phi}^{-1}/u)$ and  the emerging ``dephasing'' time is the electron lifetime $\tau_{\phi}^{-1}=\tau_F^{-1}=2\pi\gamma T$. This enables us to verify a basic property of phase breaking: the damping of the oscillation amplitude scales with the length of the interfering paths.  For weak interactions, one might also resort to the Johnson-Nyquist noise approach of Sec. III C to recover that $\tau_{\phi}^{-1}\propto T$. However, we must stress that, in our geometry, the latter is only applicable in the limit of very weak interactions
where Eq. (\ref{trente2}) becomes completely similar to Eq. (22). Here, we like to underline that interference experiments of Ref. \cite{Hansen,Kobayashi} performed in the ballistic limit report a dephasing time that varies as $1/T$. It is relevant to remember that those results might be explained from the electron-electron interaction (that can be intrinsic or induced by side-gates). 

On the other hand, for close enough tunnel constrictions $(Td/u\rightarrow 1)$, the current $I_{\Phi}$ is only suppressed as $(T/D)^{4\gamma}$. Moreover, when $Td/u\ll 1$, $I_{\Phi}\propto V^{4\gamma+1}$ or $V(a/d)^{4\gamma}$ if $\omega=V\sim u/d\gg T$ as implicitly assumed in our Ref. \cite{KLH2}.  For wires with finite lengths $\sim L$, one would expect a saturation of the power-law suppression at low voltages $V=u/L$. When $Td/u\ll 1$, the tunneling operator involving the electron backscattering also contributes to $I_{\Phi}$ in a similar way.

If the wires exhibit distinct distances between the two tunnel points, say, 
$d_1$ and $d_2$, one must add a geometrical phase $\theta=(k_{F}d_1-k_{F}d_2)$ in the cosine,  $\cos(2\pi\Phi/\Phi_o)\rightarrow \cos(k_{F}d_1-k_{F}d_2-2\pi\Phi/\Phi_o)$; again, we assume that the electron densities are approximately the same in the two wires. As a result, one must properly consider the thermal averaging on the phase-coherence of Aharonov-Bohm oscillations: $\theta(T+E)-\theta(E)=\delta k_F(T)(d_1-d_2)\approx T/(E_F a)(d_1-d_2)$. Generally, one expects that the conductance oscillations might be washed out when $|\theta(T+E)-\theta(E)|\approx 2\pi$. Assuming $| d_1-d_2 |\approx a$, thus this 
leads to $T\approx E_F$. In the present geometry ({\it i.e.}, in the absence of closed orbits in the interferometer \cite{Markus}), we infer that $T\ll E_F$ is a sufficient prerequisite for neglecting the influence of thermal averaging. 

\subsection{Chirality effects}

Now, we would like to briefly comment on chirality effects. At the edges of the quantum Hall effect,  the electron operator gets modified as $\Psi_+^{\dagger}(0,0) = (1/\sqrt{2\pi a}){\cal L}_+^{Q_+=1}(0,0)$ \cite{NoteP} and thus the electron Green's function becomes:
\begin{equation}
G_+(x,\tau) = 
 - e^{ik_F x}\frac{1}{2\pi a}\left(\frac{a\pi}{u\beta}\right)^{g^{-1}} \left(\sin\left[\frac{\pi}{\beta}\left(\tau-i\frac{x}{u}\right)\right]\right)^{-g^{-1}}.
 \end{equation}
Here, $Q_+=1$ must be interpreted as $1/g$ fractionally charged quasiparticles with charge $g$ \cite{Geller} propagating together at the same velocity thus this explicitly suppresses the exponential decay of the electron Green's function with $(x'-x)$:
\begin{equation}
G_+(x'-x, V\leq T) \approx \frac{-i}{u}\theta(x'-x)e^{i k_F(x'-x)} e^{i \frac{V}{u}(x'-x)}[T/D]^{g^{-1}-1}.
\end{equation}
Let us now compare our results with those of Ref. \cite{Geller}
concerning a similar geometry but with chiral LLs formed in the strong-antidot-coupling regime in the fractional quantum Hall regime; the upper wire is replaced by an edge state
going to the {\it right} and the lower wire is substituted by an edge state counter-going to the {\it left}. 
In this case:
\begin{equation}
X_{do}^R=-i\theta(t) \frac{1}{(2\pi a)^2} e^{i2k_F d}\left(\frac{a\pi}{u\beta}\right)^{2g^{-1}} \left(\sinh\left[\frac{\pi}{\beta}\left(t-\frac{d}{u}-i0^+\right)\right]\right)^{-g^{-1}} \left(\sinh\left[\frac{\pi}{\beta}\left(t+\frac{d}{u}-i0^+\right)\right]\right)^{-g^{-1}}.
\end{equation}
Therefore, it is straightforward to note that already for non-interacting electrons $(g=1)$, owing to the fact that the electrons of the two edges propagate in opposite directions, this will drastically affect the current $I_{\Phi}$ as:
\begin{equation}
I_{\Phi} = \frac{e^2 V}{h} \frac{2\pi Td}{u}|\Gamma_0 \Gamma_d| \cos(2\pi\Phi/\Phi_o+2k_F d)\sinh^{-1}\left[\frac{\pi T 2d}{u}\right].
\end{equation}
For well-separated constrictions such that $Td/u\gg 1$, this results in an exponential attenuation of the
Aharonov-Bohm oscillations due to backscattering effects; the latter must be clearly distinguished from the electron fractionalization phenomenon described in Sec. III B. For chiral LLs $(g\neq 1)$ and $V\ll u/d\ll T$ we check that the dephasing time reads $\tau_{\phi}^{-1}= \pi T g^{-1}$ \cite{Geller}. Finally, note that for $(V,T)\ll u/d$ the conductance $G_{\Phi}$ gets only suppressed as $[\max(V,T)]^{2g^{-1}-2}$.

\section{Spinful electrons}

Now, we extend the discussion to fermions with spin that corresponds to the situation without 
Zeeman effect. Here, the electron spectrum is subject to chiral and spin-charge decomposition.
Thus, one expects two types of excitations, namely the {\it fractional charges} described above and 
 {\it spinons} (the spin-1/2 excitations of purely low-dimensional quantum spin systems). Below, we seek to revisit the electron lifetime in the presence of spin-charge separation.

\subsection{Spin-Charge separation}

To estimate the electron lifetime, we find it  appropriate to exploit the precise decomposition (consult Appendix B)
\begin{equation}
\Psi_{+\uparrow}^{\dagger}(0,0) 
=\frac{1}{\sqrt{2\pi a}}\left({\cal C}_+^{\frac{1+g}{2}}{\cal C}_-^{\frac{1-g}{2}}
{\cal S}_+^{1}\right)(0,0).
\end{equation}
Following the same procedure as for spinless electrons we derive the electron Green's function
\begin{eqnarray}
G_{+\uparrow}(x,\tau>0) &=& 
-e^{i k_F x}\langle T_{\tau} \Psi_{+\uparrow}(x,\tau)\Psi^{\dagger}_{+\uparrow}(0,0)\rangle \\ \nonumber
&=& - e^{ik_F x}\frac{a^{\gamma}}{2\pi}\left(\frac{\pi}{u\beta}\right)^{\gamma+1} \left(\sin\left[\frac{\pi}{\beta}\left(\tau-i\frac{x}{u}\right)\right]\right)^{-\frac{\gamma+1}{2}}\left(\sin\left[\frac{\pi}{\beta}\left(\tau+i\frac{x}{u}\right)\right]\right)^{-\frac{\gamma}{2}} \sin^{-\frac{1}{2}}\left[\frac{\pi}{\beta}\left(\tau-i\frac{x}{v_F}\right)\right].
\end{eqnarray}
Hence, we estimate
\begin{equation}
G_{+\uparrow}(x'-x,E\rightarrow 0,T) \approx \frac{-i}{u}\theta(x'-x)e^{i k_F(x'-x)} [T/D]^{\gamma} \exp\left[-\frac{(x'-x)}{u{\tau}_F}\right],
\end{equation}
where
\begin{equation}
\label{spin}
{\tau}_F^{-1}=\pi T\left(\gamma+\frac{1}{2}\left(\frac{u}{v_F}-1\right)\right).
\end{equation}
The first part in Eq. (\ref{spin}) stems from the fractionalization in the charge sector whereas the second part emphasizes the difference between the spin and the charge thermal lengths, $\xi_s \approx \beta v_F$ and $\xi_c\approx \beta u$; assuming $\gamma$ finite, $\xi_c$ can be identified as the typical length scale at which the electron loses any physical meaning due to the chiral decomposition in the charge sector (consult Fig. 3) whereas $\xi_s$ embodies the typical spin diffusion length. Note that the bosonization allows us to compute the electron Green's function in a non-perturbative manner and to conclude that spin-charge separation engenders an extra contribution $\pi T(-1+u/v_F)$ in $\tau_F^{-1}$ that already dominates over the chiral decomposition for weak interactions; as a result, in agreement with Ref. \cite{Gornyi}, we identify $\tau_F^{-1}\propto T(Ua/v_F)$. Another difference with Eq. (\ref{density}) is that the tunneling density of states is proportional $T^{\gamma}$ (instead of $T^{2\gamma}$). Concerning the interference geometry of Fig. 5, one expects a factor $T^{2\gamma}$ (instead of $T^{4\gamma}$) in the flux-dependent piece of the current in Eq. (\ref{AB}). 

Let us stress that the perturbative approach of Sec. II could not be applied to
spinful electrons because the forward scattering diagrams of Fig. 1 would not cancel each other. The (unphysical) strong-delta-function singularity on the mass shell due to second-order diagrams does not vanish \cite{Gornyi}. For spinful electrons, scattering from the same chiral branch must be treated in a non-perturbative manner to reproduce the difference between spin and charge velocities.

The usual LL assumes interacting electrons in which the interaction strength is not too great resulting in a spin velocity $v_s=v_F$. On the other hand, at low electron densities the potential energy grows relative to the kinetic energy and eventually dominates it for sufficiently low densities when $na_B\ll 1$, with $n$ being the average density of the electrons and the Bohr radius 
$a_B=\epsilon/m$, with $\epsilon$ the dielectric constant and $m$ the mass of the electron (to be consistent with the text we have taken $e=1=\hbar$). At these low densities, where the system can be viewed as a fluctuating Wigner solid \cite{M1}, there is a natural separation of
energy scales between the magnetic exchange energy \cite{Fogler,Klironomos}, $J\sim D e^{(-1/\sqrt{n a_B})}$ and the plasmon (phonon) energy $D\sim u/a$ with the renormalized lattice spacing $a=n^{-1}$
for decreasing densities.  When the interactions between electrons become very strong, they must tunnel through one another to exchange, leading to the myriad of energy scales $J \ll D$ or $v_s=Ja\ll u$. Therefore, we expect that the electron lifetime of Eq. (\ref{spin}) will be modified accordingly as ${\tau}_F^{-1}\rightarrow \pi Tu/(2 v_s)$. In the interference geometry considered in Fig. 5, we thus expect a suppression of the Aharonov-Bohm oscillations when the distance $d$ between the tunneling positions will exceed $\xi_s\sim v_s/T (\ll \xi_c\sim u/T)$ that tends to the lattice spacing $a$ when $T\rightarrow J^-$. {\it It is important to bear in mind that in the low-density regime and in the limit $\xi_s\ll d\ll \xi_c$ electron interferences are irrefutably suppressed as a result of the spin excitations that cannot propagate through the interference region}.

The form factor for the momentum-resolved density of states will be broadened by an amount of order $\delta k = (u\tau_F)^{-1}\rightarrow T/v_s = \xi_s^{-1}$. In the low-density realm, assuming $J\geq T$, the spectral function for spinful electrons will exhibit peaks centered at $k_F=\pi n/2$ that
become strongly broadened when $T\rightarrow J$ due to spin-charge separation.

The LL considered above implies that $T\ll J$ or $\xi_s\gg a$ (allowing us to use the continuum limit for spin excitations). Now, one might wonder how the results will be modified in the magnetically-incoherent realm in which $J\leq T$ \cite{FLB}.

\subsection{Crossover to the spin-incoherent LL}

 The condition $J<T$ means that the spin degrees of freedom become nondynamical in that, within the ``thermal coherence time''  $1/T$, the spin quantum numbers of individual electrons remain unchanged, since a spin-flip transition requires a time $1/J$ to occur. Hence, dynamically, the electron gas completely behaves in a ``spinless'' fashion, since the spin degrees of freedom are static and random, and do not couple to the electron coordinates anymore. As a result, one naturally expects the doubling of the Fermi momentum when $T<J$ leading to the rescaling $k_F\rightarrow \pi n$.

More precisely, the electron Green's function may be explicitly expressed as \cite{Balents,Fiete}
\begin{equation}
G_{+\uparrow}(x'-x,\tau) \sim \langle 2^{-|\hat{N}(x'-x,\tau)|} \Psi(x'-x,\tau) \Psi^{\dagger}(0,0) \rangle,
\end{equation}
where we have introduced the spinless fermion formalism of Appendix A and $\hat{N}(x'-x,\tau)$ embodies the particle number (operator) between points $x$ and $x'$. Since the spin dynamics is effectively frozen out (exchange events do not occur) all electrons between the points $x$ and $x'$ must have parallel spins to contribute to the Green's function --- clarifying the factor $2^{-|\hat{N}|}$ --- that is unlikely. Thus, at relatively large distances, this explicitly results in:
\begin{equation}
G_{+\uparrow}(x'-x,\tau) \propto 2^{-n|x'-x|}=\exp(-n|x'-x|\ln 2).
\end{equation}
In the spin-incoherent realm, the spin coherence length $\xi_s$, at which the exponential decay is significant, can be precisely identified as $(k_F\ln 2/\pi)^{-1}$, where the Fermi momentum now obeys $k_F=\pi n=\pi/a$ reflecting the spinless aspect of the propagating particles in the magnetically-incoherent regime. This already leads to the important conclusion that the momentum resolved-density of states now yields broad peaks centered at $\pm \pi n$ (instead of $\pi n/2$ for $T\leq J$) if the Wigner crystal is quite robust or if the root-mean-square displacement of an electron is smaller than $a$ \cite{Fiete}.

Now, let us discuss the Aharonov-Bohm setup of Fig. 5 in the spin-incoherent realm. To compute explicitly the electron Green's function $G_{+\uparrow}(x'-x,\tau)$ one needs to notice two important points. First, the mapping to the spinless LL requires the proper rescaling of the Luttinger parameter as $g^*=2g$ \cite{FLB}: keep in mind that in the spin-incoherent regime, $g^*$ embodies the Luttinger parameter of the effective spinless theory, reflecting that real electrons now must behave as weakly-interacting spinless fermions close to $g=1/2$. Second, the fluctuating piece in $\hat{N}(x'-x,\tau)$ will result in a small anomalous term in $\gamma$: indeed, $\gamma=(Q_-^*)^2/g^* - g^*\left(\ln 2/\pi\right)^2/4$ and by analogy with the spinless case we derive $Q_-^*=(1-g^*)/2$. The electron Green's function can be found explicitly in Refs. \cite{Cheianov,Balents,Fiete}. This allows us to compute the Aharonov-Bohm type conductance $G_{\Phi}$ thoroughly and by assuming $(T,V)\rightarrow 0$ we obtain
\begin{equation}
G_{\Phi} \propto \frac{e^2}{h}[\max(V,T)/D]^{\frac{1}{g^*}+g^*\left[1-\left(\frac{\ln 2}{\pi}\right)^2\right]-2} e^{-2d\ln 2/a}\sqrt{\Gamma_0 \Gamma_d} \cos(2\pi\Phi/\Phi_o).
\end{equation}
We observe a strong exponential reduction as soon as $d\gg a=n^{-1}$ reflecting that the spin coherence length is very small in the magnetically-incoherent regime because the spins are static and random. Let us mention that  Eq. (57) is in accordance with the recent results found in Ref. \cite{Kindermann}.  Note that the exponential reduction for $d\gg a$ also agrees with the results obtained for $T\rightarrow J^-$. On the other hand, for very small distances between the tunneling points O and P, {\it i.e.}, $d\leq a$, one may observe a visible conductance $G_{\Phi}$, even when $(T,V)\rightarrow 0$ if $g$ is close enough to $1/2$. 

\section{Conclusion}

To summarize, we have studied the decoherence of the electron wavepacket in LLs thoroughly and
fleshed out the results of our previous Refs. \cite{KLH1,KLH2}. We have elucidated that, for long times such that $2\pi T t\geq 1$, the electron Green's function $G_{\pm}(p,t)$ always falls off exponentially with time and the corresponding electron lifetime varies as $T^{-1}$ for spinful and spinless electrons. The electron lifetime cuts off the power-law decay with time of the electron Green's function in the quantum realm. We have provided a physical justification of the electron lifetime as well of the anomalous mass effects close to the Fermi level in terms of the fractionalization of the electron wavepacket. For strong interactions, we verify that  $(T\tau_F)\ll 1$, reflecting that in LLs the electron is not a good Landau quasiparticle. In Table 1, we have listed the electron lifetime as well as the emergent fractional excitations in the different situations. 

\vskip 0.5cm

\begin{center}
\begin{tabular}{||c|c|c|c|c||}
\tableline
Regime & Fractional excitations & $\tau_F^{-1}$ \\
\tableline

spinless & $Q_{\pm}=(1\pm g)/2$ & $2\pi [(Q_-)^2/g] T$  \\

$g\rightarrow 1$ &  & $\sim (U/E_F)^2 T$\\

\tableline
spinful &  $Q_{\pm}=(1\pm g)/2$ and $|S_z|=1/2$ $(\xi_s=v_s/T)$ & $\pi T[(Q_-)^2/g+(-1+u/v_s)]$ \\

$g\rightarrow 1$ & & $\sim (U/E_F)T$ \\

\tableline
spin-incoherent  $J\ll T$ &  $Q_{\pm}^*=(1 \pm g^*)/2$ and $\xi_s=a/\ln 2$ &  $\sim u/\xi_s$ \\
\tableline
\end{tabular}
\end{center}

\vskip 0.05cm
{\it Table 1:  Fractional excitations and the electron lifetime for spin-polarized and spinful electrons.   The injected electron carries a wavevector at $+k_F$; for an electron close
to $-k_F$ one must simply exchange the role of $Q_+$ and $Q_-$.}
\vskip 0.5cm

For weak interactions and spin-polarized electrons, we obtain $\tau_F^{-1} \propto (U/E_F)^2 T$ in accordance with the Hartree-Fock diagrammatic approach \cite{Chubukov,Maslov}. For spinful electrons,  spin-charge separation substantially enhances the decoherence of the electron wavepacket and for weak interactions we find $\tau_F^{-1} \propto (U/E_F)T$ in agreement with Ref. \cite{Gornyi}. For strong interactions or in the low-density limit, spin excitations can hardly propagate decreasing considerably the electron lifetime $(\tau_F^{-1} \sim Tu/v_s)$. For time scales $\sim \tau_F$, the charge excitations still move coherently. In the spin-incoherent realm where the spin exchange energy $J$ becomes smaller than $T$, the spin state of the conductor becomes static and random, and the spin diffusion length $\xi_s$ fatally converges to the lattice spacing $a$ or $\tau_F^{-1}\sim u\ln 2/a$, {\it i.e.}, $\xi_s\ll \xi_c$.

We like to emphasize that the exponential decay with time or space of the electron Green's function, allowing to reveal the electron lifetime, can be probed experimentally through momentum resolved tunneling or Aharonov-Bohm type measurements via two tunnel-coupled wires. At finite temperature, the momentum resolved tunneling density of states exhibits a Lorentzian type profile centered at $\pm k_F$ and the broadening must be identified as $\delta k=\frac{1}{u}\tau_F^{-1}$. For spinless electrons, 
corresponding to a large Zeeman effect, the peaks are quite sharp since the broadening can never exceed ${\cal O}(T/v_F)$. On the other hand, for spinful electrons, the broadening can reach $\xi_s^{-1}\sim k_F$ in the low-density magnetically-incoherent realm $(J\leq T)$ emphasizing that the
physical electron is certainly not a good Landau quasiparticle. Another interesting feature that might be verified experimentally in the low density regime is the doubling of the Fermi wavevector when entering into the spin-incoherent limit, reflecting that propagating particles behave as (weakly-interacting) spinless impenetrable fermions; remember, the spins do not couple to the electrons coordinates anymore. Moreover, we have carefully demonstrated that the electron lifetime can be identified as the dephasing time in an Aharonov-Bohm type geometry involving two weakly tunnel-coupled (very long) wires. We like to stress that a dephasing time varying as $T^{-1}$ has been reported in Refs. \cite{Hansen,Kobayashi} that seems to reinforce the thesis that electron-electron interactions should not be ignored in quantum wires. The work presented here can be extended in many directions: in particular, muti-channels and systems with backscattering may be considered. Finally, note that for weak-localization type experiments the dephasing time is not necessarily the electron lifetime \cite{Gornyi}.

\acknowledgements
We acknowledge very useful discussions with L. Balents, M. B\"{u}ttiker, G. Fiete, B. Halperin, A. Mirlin, K.-V. Pham, D. Polyakov, M. Randeria, and I. Safi. This work was supported by CIAR, FQRNT, and NSERC.

\begin{appendix}

\section{Electron Green's function}

\subsection{Bosonization}

The electron operator can be decomposed into $\Psi(x)=e^{i k_F x}\Psi_+(x) + e^{-i k_F x} \Psi_-(x)$; the right (+) and left (-) moving fermions are decsribed through the annihilation operators:
\begin{eqnarray}
\Psi_+(x) &=&\frac{1}{\sqrt{2\pi a}}: \exp i\sqrt{\pi}(\theta(x)-\phi(x)): \\ \nonumber
\Psi_-(x) &=& \frac{1}{\sqrt{2\pi a}}: \exp i\sqrt{\pi}(\theta(x)+\phi(x)):.
\end{eqnarray}
These operators are assumed to be normal ordered. The Fermi momentum $k_F=\pi N/L$ is fixed
by the number of particles $N$ or the chemical potential. Hence, we have the precise identifications
\begin{eqnarray}
\hat{N} &=& N+\hat{Q} = N-\frac{1}{\sqrt{\pi}}\int_0^L \nabla \phi(x) dx, \\ \nonumber
\hat{J} &=& \frac{1}{\sqrt{\pi}}\int_0^L \nabla \theta(x) dx.
\end{eqnarray}
The charge operator $\hat{Q}=N_+ +N_-$ --- with $N_+$ and $N_-$
being the (integral) number of electrons added to the ground state at the right and left Fermi points --- has integral values; the charge $Q=\langle \hat{Q}\rangle$ has been normalized to the charge of the electron. Moreover, we identify $\hat{J}=N_+-N_-$ implying that the current  has been normalized to the Fermi velocity $v_F$. In one dimension, the interacting fermionic Hamiltonian is equivalent to a gaussian bosonic model \cite{Giamarchi}
\begin{equation}
\label{Lut}
{\cal H}=\frac{u}{2}\int_{0}^{L} dx\  \frac{1}{g}\left(\partial_x\phi\right)^2+g\left(\partial_x\theta\right)^2,
\end{equation}
where $\theta$ and $\nabla\phi$ are canonically conjugate
\begin{equation}
[\theta(x),\nabla \phi(y)]=i\delta(x-y).
\end{equation}
The parameter $u=v_F/g$ embodies the boson (plasmon) velocity and $g<1$ measures the strength of the interactions. For free electrons, $u=v_F$ and $g=1$. The forward $(g_4)$ and backward $(g_2)$ scatterings between electrons lead to:
\begin{eqnarray}
\label{ug}
u &=& \sqrt{\left(v_F+\frac{g_4}{2\pi}\right)^2-\left(\frac{g_2}{2\pi}\right)^2},\\  \nonumber
g &=& \sqrt{\frac{2\pi v_F+g_4-g_2}{2\pi v_F+g_4+g_2}},
\end{eqnarray}
with $g_2=g_4=(Ua)$, $U$ being the repulsion between electrons. Below, we use the chiral decomposition \cite{KV} 
\begin{equation}
\theta_{\pm}=\theta\mp\phi/g
\end{equation}
such that the Hamiltonian turns explicitly into
\begin{equation}
{\cal H}=H_+ +H_-=\frac{ug}{4}\int_0^L dx\ (\partial_x\theta_{+})^2+(\partial_x\theta_{-})^2.
\end{equation}
We check that $[H_+,H_-]=0$ and the equations of motion for those fields are $u\partial_x\theta_{\pm}=\mp\partial_t\theta_{\pm}$, thus $\theta_{\pm}(x,t)=\theta_{\pm}(x\mp ut)$.

{\it Now, we provide a pedestrian derivation of the electron Green's function in the LL at finite temperature. }

\subsection{Free electrons}

We begin with free electrons $(g=1)$. For convenience, we redefine the electron operator $\Psi_+\rightarrow \Psi_+ \sqrt{2\pi a}$ such that $\Psi_+=:\exp(i\sqrt{\pi}\theta_+):$ with the chiral field
$\theta_+=\theta-\phi$. Thus, exploiting the bosonic theory, we identify
\begin{eqnarray}
<\Psi_+(x,\tau)\Psi_+^{\dagger}(0,0)>\ &=& <\exp[i\sqrt{\pi}\theta_+(x,\tau)]
\exp[-i\sqrt{\pi}\theta_+(0,0)]>\\ \nonumber
&=& \exp[\pi<\theta_+(x,\tau)\theta_+(0,0)-\theta_+(x,\tau)^2>].
\end{eqnarray}
Having in mind that we have redefined $\Psi_+\rightarrow \Psi_+ \sqrt{2\pi a}$ and, {\it assuming $\tau>0$}, a direct calculation gives
\begin{equation}
<\Psi_+(x,\tau)\Psi^{\dagger}_+(0,0)>\ =\left(\frac{a\pi/(v_F\beta)}{
\sin(\frac{\pi}{\beta}[\tau-i\frac{x}{v_F}])}\right),
\end{equation}
resulting in the precious formulas
\begin{equation}
<\theta_+(x,\tau)\theta_+(0,0)-\theta_+(x,\tau)^2>\ =\frac{1}{\pi}\ln\left[
\frac{a\pi/(v_F\beta)}{\sin(\frac{\pi}{\beta}[\tau-i\frac{x}{v_F}])}\right].
\end{equation}
Similarly, using left-moving electron operators, we identify
\begin{equation}
<\theta_-(x,\tau)\theta_-(0,0)-\theta_-(x,\tau)^2>\ =\frac{1}{\pi}\ln\left[
\frac{a\pi/(v_F\beta)}{
\sin(\frac{\pi}{\beta}[\tau+i\frac{x}{v_F}])}\right].
\end{equation}
Using the definitions of $\theta_{\pm}$, then we reach
\begin{eqnarray}
\label{one}
<\phi(x,\tau)\phi(0,0)&-&\phi(0,0)\phi(0,0)+
\theta(x,\tau)\theta(0,0)
-\theta(0,0)\theta(0,0)> \\ \nonumber
&=& \frac{1}{\pi}\ln\left[
\frac{a\pi}{(v_F\beta)}
\frac{1}{[\sin(\frac{\pi}{\beta}[\tau+i\frac{x}{v_F}])]^{1/2}
[\sin(\frac{\pi}{\beta}[\tau-i\frac{x}{v_F}])]^{1/2}}\right],
\end{eqnarray}
and,
\begin{eqnarray}
\label{two}
<\phi(x,\tau)\theta(0,0)&-&\phi(0,0)\theta(0,0)+
\theta(x,\tau)\phi(0,0)
-\theta(0,0)\phi(0,0)> \\ \nonumber
&=&\frac{1}{\pi}\ln\left[
\frac{\sin(\frac{\pi}{\beta}[\tau+i\frac{x}{v_F}])^{1/2}}
{\sin(\frac{\pi}{\beta}[\tau-i\frac{x}{v_F}])^{1/2}}\right].
\end{eqnarray}

\subsection{Interacting electrons}

Interacting electrons are embodied by the Luttinger Hamiltonian in Eq. (\ref{Lut}). To compute the electron Green's function with electron-electron interactions, we redefine $\phi=\sqrt{g}\hat{\phi}$ and $\theta=\hat{\theta}/\sqrt{g}$ such
that
\begin{equation}
{\cal H}=\frac{u}{2}\int dx\ (\partial_x\hat{\phi})^2+(\partial_x\hat{\theta})^2.
\end{equation}
This Hamiltonian is equivalent to that of fictitious free electrons $\hat{\Psi}$ if we replace $v_F$ by $u$ and make, {\it e.g.}, the identification $\hat{\Psi}_+ = \exp i\sqrt{\pi}(\hat{\theta}-\hat{\phi})$. By identification with the bosonic formulas ({\ref{one}}) and ({\ref{two}}), we deduce
\begin{eqnarray}
<\hat{\phi}(x,\tau)\hat{\phi}(0,0)-\hat{\phi}(0,0)\hat{\phi}(0,0)+
\hat{\theta}(x,\tau)\hat{\theta}(0,0)
-\hat{\theta}(0,0)\hat{\theta}(0,0)> \\ \nonumber
= \frac{1}{\pi}\ln\left[
\frac{a\pi}{(u\beta)}\frac{1}{[\sin(\frac{\pi}{\beta}[\tau+i\frac{x}{u}])]^{1/2}
[\sin(\frac{\pi}{\beta}[\tau-i\frac{x}{u}])]^{1/2}}\right].
\end{eqnarray}
By symmetry between the fields $\hat{\phi}$ and $\hat{\theta}$ in the 
Hamiltonian, we also infer
\begin{eqnarray}
<\hat{\phi}(x,\tau)\hat{\phi}(0,0)-\hat{\phi}(0,0)\hat{\phi}(0,0)>\ =\ <\hat{\theta}(x,\tau)\hat{\theta}(0,0)-\hat{\theta}(0,0)\hat{\theta}(0,0)>\\ \nonumber
=\frac{1}{2\pi}\ln\left[
\frac{a\pi}{(u\beta)}\frac{1}{[\sin(\frac{\pi}{\beta}[\tau+i\frac{x}{u}])]^{1/2}
[\sin(\frac{\pi}{\beta}[\tau-i\frac{x}{u}])]^{1/2}}\right].
\end{eqnarray}
Additionally, we have
\begin{eqnarray}
<\hat{\phi}(x,\tau)\hat{\theta}(0,0)-\hat{\phi}(0,0)\hat{\theta}(0,0)+
\hat{\theta}(x,\tau)\hat{\phi}(0,0)
-\hat{\theta}(0,0)\hat{\phi}(0,0)> \\ \nonumber
=\frac{1}{\pi}\ln\left[
\frac{\sin(\frac{\pi}{\beta}[\tau+i\frac{x}{u}])^{1/2}}
{\sin(\frac{\pi}{\beta}[\tau-i\frac{x}{u}])^{1/2}}\right].
\end{eqnarray}
Now, exploiting the precious identification $\Psi_+=\exp[i\sqrt{\pi}(-\sqrt{g}\hat{\phi}+\frac{1}{
\sqrt{g}}\hat{\theta})]$ we obtain
\begin{eqnarray}
<\Psi_+(x,\tau)\Psi_+^{\dagger}(0,0)>
&=& <\exp[i\sqrt{\pi}(-\sqrt{g}\hat{\phi}+\frac{1}{
\sqrt{g}}\hat{\theta})(x,\tau)]
\exp[-i\sqrt{\pi}(-\sqrt{g}\hat{\phi}+\frac{1}{
\sqrt{g}}\hat{\theta})(0,0)]>\\ \nonumber
&=&\exp\{\pi[g<\hat{\phi}(x,\tau)\hat{\phi}(0,0)-\hat{\phi}(0,0)\hat{\phi}(0,0)>+\frac{1}{g}<\hat{\theta}(x,\tau)\hat{\theta}(0,0)-\hat{\theta}(0,0)
\hat{\theta}(0,0)>\\ \nonumber
&+&<
\hat{\phi}(x,\tau)\hat{\theta}(0,0)-\hat{\phi}(0,0)\hat{\theta}(0,0)+
\hat{\theta}(x,\tau)\hat{\phi}(0,0)
-\hat{\theta}(0,0)\hat{\phi}(0,0)>]\},
\end{eqnarray}
resulting in
\begin{eqnarray}
\hskip -0.3cm
<\Psi_+(x,\tau)\Psi_+^{\dagger}(0,0)>\ =
{\left[\frac{a\pi}{(u\beta)}\right]}^{\frac{g+g^{-1}}{2}}
\left[\sin\left(\frac{\pi}{\beta}[\tau+i\frac{x}{u}]\right)\right]^{-\frac{g+g^{-1}}{4}}
\left[\sin\left(\frac{\pi}{\beta}[\tau-i\frac{x}{u}]\right)\right]^{-\frac{g+g^{-1}}{4}}
\frac{\sin(\frac{\pi}{\beta}[\tau+i\frac{x}{u}])^{1/2}}
{\sin(\frac{\pi}{\beta}[\tau-i\frac{x}{u}])^{1/2}},
\end{eqnarray}
and finally in
\begin{equation}
<\Psi_+(x,\tau)\Psi_+^{\dagger}(0,0)>\ = a^{2\gamma+1}
{\left[\frac{\pi}{u\beta}\right]}^{2\gamma+1}
\left[\sin\left(\frac{\pi}{\beta}\left[\tau+i\frac{x}{u}\right]\right)\right]^{-\gamma}
\left[\sin\left(\frac{\pi}{\beta}\left[\tau-i\frac{x}{u}\right]\right)\right]^{-\gamma-1},
\end{equation}
where
\begin{equation}
\gamma=\frac{(g+g^{-1})}{4}-1/2.
\end{equation}
We can return to the physical fermions by shifting $\Psi_+\rightarrow \Psi_+/\sqrt{2\pi a}$ resulting in
\begin{equation}
<\Psi_+(x,\tau)\Psi_+^{\dagger}(0,0)>\ = \frac{a^{2\gamma}}{2\pi}
{\left[\frac{\pi}{u\beta}\right]}^{2\gamma+1}
\left[\sin\left(\frac{\pi}{\beta}\left[\tau+i\frac{x}{u}\right]\right)\right]^{-\gamma}
\left[\sin\left(\frac{\pi}{\beta}\left[\tau-i\frac{x}{u}\right]\right)\right]^{-\gamma-1}.
\end{equation}
Applying a Wick rotation in time space we finally obtain the Green's function ($t>0$ has been assumed)
 \begin{equation}
<\Psi_+(x,t)\Psi_+^{\dagger}(0,0)>\ \approx -i\frac{a^{2\gamma}}{2\pi}
{\left[\frac{\pi}{u\beta}\right]}^{2\gamma+1}
\left[\sinh\left(\frac{\pi}{\beta}[t+\frac{x}{u}-i0^+]\right)\right]^{-\gamma}
\left[\sinh\left(\frac{\pi}{\beta}[t-\frac{x}{u}-i0^+]\right)\right]^{-\gamma-1},
\end{equation}
where $0^+$ is a short-time cutoff. This result agrees, {\it e.g.}, with that of Ref. \cite{Gornyi} if one identifies $\Lambda\approx u/a \approx 1/0^+$ such that there is a unique ultraviolet cutoff. 
In the quantum limit of $T=0$,  we obtain 
 \begin{eqnarray}
<\Psi_+(x,t)\Psi_+^{\dagger}(0,0)> &=& -i\frac{a^{2\gamma}}{2\pi}
\frac{1}{{u}^{2\gamma+1}}\left[t+\frac{x}{u}-i0^+\right]^{-\gamma}
\left[t-\frac{x}{u}-i0^+\right]^{-\gamma-1} \\ \nonumber
&=& -i \frac{a^{2\gamma}}{2\pi} \left[ut+x-i0^+\right]^{-\gamma}
\left[ut-x-i0^+\right]^{-\gamma-1}.
\end{eqnarray}
Assuming $t>0$, we infer the time-ordered electron Green's function
\begin{equation}
\label{TGreen}
G_+(x,t>0)= -i e^{i k_F x} <T\Psi_+(x,t)\Psi_+^{\dagger}(0,0)>\ = e^{i k_F x} \frac{a^{2\gamma}}{2\pi}\frac{1}{x-u t+i0^+}
\left[\left(x-ut+i0^+\right)\left(x+ut-i0^+\right)\right]^{-\gamma}.
\end{equation}

\subsection{Bosonic correlators}

Here, we extract the propagators for the chiral boson fields $\theta_{\pm}=(\theta\mp \phi/g)$. 
Using the results above, we infer
\begin{eqnarray}
\label{phi}
<{\phi}(x,\tau){\phi}(0,0)-{\phi}(0,0){\phi}(0,0)>\ =\ \frac{g}{2\pi}\ln\left[
\frac{a\pi}{(u\beta)}\frac{1}{[\sin(\frac{\pi}{\beta}[\tau+i\frac{x}{u}])]^{1/2}
[\sin\left(\frac{\pi}{\beta}[\tau-i\frac{x}{u}]\right)]^{1/2}}\right],
\end{eqnarray}
as well as
\begin{eqnarray}
<{\theta}(x,\tau){\theta}(0,0)-{\theta}(0,0){\theta}(0,0)>\ =\ \frac{1}{2g\pi}\ln\left[
\frac{a\pi}{(u\beta)}\frac{1}{[\sin(\frac{\pi}{\beta}[\tau+i\frac{x}{u}])]^{1/2}
[\sin(\frac{\pi}{\beta}[\tau-i\frac{x}{u}])]^{1/2}}\right],
\end{eqnarray}
and finally 
\begin{eqnarray}
\label{theta}
<{\phi}(x,\tau){\theta}(0,0)-{\phi}(0,0){\theta}(0,0)+
{\theta}(x,\tau){\phi}(0,0)
-{\theta}(0,0){\phi}(0,0)> \\ \nonumber
=\frac{1}{\pi}\ln\left[
\frac{\sin(\frac{\pi}{\beta}[\tau+i\frac{x}{u}])^{1/2}}
{\sin(\frac{\pi}{\beta}[\tau-i\frac{x}{u}])^{1/2}}\right].
\end{eqnarray}
In terms of the chiral fields $\theta_{\pm}=\theta\mp \phi/g$ this leads to the important bosonic propagators
\begin{equation}
<\theta_{\pm}(x,\tau)\theta_{\pm}(0,0)-\theta_{\pm}(x,\tau)^2>\ 
=\frac{1}{g\pi}\ln\left(
\frac{a\pi/(u\beta)}{\sin(\frac{\pi}{\beta}[\tau\mp i\frac{x}{u}])}\right).
\end{equation}
We can check that $\theta_{\pm}(x,\tau)=\theta_{\pm}(x\pm iu\tau)$. Note that Eq. ({\ref{phi}}) can be also
rewritten as \cite{Giamarchi}
\begin{eqnarray}
-2<{\phi}(x,\tau){\phi}(0,0)-{\phi}(0,0){\phi}(0,0)>\ =\ \frac{g}{\pi}\ln\left[\left(\frac{u\beta}{a\pi}\right)^2 \left(\sinh^2\left(\frac{\pi x}{\beta u}\right)+\sin^2\left(\frac{\pi\tau}{\beta}\right)\right)\right].
\end{eqnarray}
Eq. ({\ref{theta}}) can be manipulated as follows. We introduce a phase $\eta$ such that $\cos\eta=\tan(\pi\tau/\beta)$ and $\sin\eta=\tanh(\pi x/\beta u)$: in the thermodynamical 
$(x\rightarrow +\infty)$ and low-energy $(\tau\rightarrow \beta)$ limit, we check $\cos^2\eta+\sin^2\eta=1$. Hence
\begin{eqnarray}
e^{i\eta} &=& \tan\frac{\pi\tau}{\beta} + i \tanh \frac{\pi x}{\beta u} 
= \frac{\sin(\frac{\pi}{\beta}[\tau+i\frac{x}{u}])}{\cos\left(\frac{\pi\tau}{\beta}\right)
\cos\left(\frac{\pi}{\beta}\frac{i x}{u}\right)},\\ \nonumber
e^{-i\eta} &=& \tan\frac{\pi\tau}{\beta} - i \tanh \frac{\pi x}{\beta u} 
= \frac{\sin(\frac{\pi}{\beta}[\tau-i\frac{x}{u}])}{\cos\left(\frac{\pi\tau}{\beta}\right)
\cos\left(\frac{\pi}{\beta}\frac{i x}{u}\right)}.
\end{eqnarray}
We infer
\begin{equation}
2i\arctan\left[\frac{\tanh(\pi x/\beta u)}{\tan(\pi\tau/\beta)}\right]=2i\eta=
\ln\left(\frac{e^{i\eta}}{e^{-i\eta}}\right)=\ln\left[\frac{\sin(\frac{\pi}{\beta}[\tau+i\frac{x}{u}])}{\sin(\frac{\pi}{\beta}[\tau-i\frac{x}{u}])}\right],
\end{equation}
and finally \cite{Giamarchi}
\begin{eqnarray}
<{\phi}(x,\tau){\theta}(0,0)-{\phi}(0,0){\theta}(0,0)+
{\theta}(x,\tau){\phi}(0,0)
-{\theta}(0,0){\phi}(0,0)>\  
=\frac{i}{\pi} \arctan\left[\frac{\tanh(\pi x/\beta u)}{\tan(\pi\tau/\beta)}\right].
\end{eqnarray}

\section{On fractionalization}

\subsection{Spin-polarized electrons}

Let us inject a charge $Q=\langle \hat{Q}\rangle$ and a current $J=\langle \hat{J} \rangle$ above the ground state.  At a first glance, one can introduce two counterpropagating states with arbitrary charges $Q_+$ and $Q_-$ \cite{Safi,KV}. Since the plasmon velocity $u$ is the only relevant velocity in the
LL, hence one can derive the precious conservation laws (charge and current conservations)
\begin{eqnarray}
Q &=& Q_+ + Q_- \\ \nonumber
v_F  J &=& u(Q_+ - Q_-).
\end{eqnarray}
From Eq. (\ref{ug}) with $g_4=g_2$ we observe that $ug=v_F$. Therefore, those conservation laws strongly suggest the emergence of fractional excitations with charges $Q_{\pm}=(Q\pm gJ)/2$ in the LL. Notice that for free electrons $(g=1)$, if one injects a particle at the right Fermi point implying $Q=J=1$, thus one recovers the physical result that $Q_+=1$ and $Q_-=0$. Taking formally the extreme limit $g\rightarrow 0$ we identify symmetric counterpropagating charge wavepackets (kinks) $Q_{\pm}=1/2$ which are reminiscent of the spinons (domain walls) in the Heisenberg chain. Moreover, following Ref. \cite{KV}, one can rigorously prove that the Fourier transforms of the $Q_{\pm}$ fractional quasiparticle operators given by (the lowerscript $\pm$ refers to the direction of propagation)
\begin{eqnarray}
{\cal L}_{\pm}^{Q_{\pm}}(x,t) &=& \exp [-i\sqrt{\pi}Q_{\pm}\theta_{\pm}(x,t)],\\ \nonumber
 &=&\exp\left[-i\sqrt{\pi}\left(\frac{Q\pm gJ}{2}\right)\left(\theta\mp \phi/g\right)(x,t)\right],
\end{eqnarray}
are exact eigenstates of the Luttinger Hamiltonian with $Q\neq 0$ and $J\neq 0$. We stress that those fractional excitations with charges $Q_{\pm}=(Q\pm gJ)/2$ are the genuine Landau quasiparticles in LLs associated with a change in the number of electrons or current. For pure current processes $(Q=0)$ one can  build a Laughlin type  wavefunction for those fractional quasiparticles \cite{KV} by analogy to the edge states of the fractional quantum Hall effect \cite{Stone}. {\it Fractional quasiparticles must be distinguished from usual plasmon excitations (particle-hole pairs) that rather conserve the number of electrons}. 

Note in passing that going around a ring in the LL we get a (persistent) current which must be quantized: ${\cal J}_R=u(Q_+-Q_-)=ugJ$. For a very clean LL $(ug=v_F)$ we obtain ${\cal J}_R=v_F J$ whereas
in the presence of umklapp scatterings or disorder one might observe a strong reduction of the current or of the product $ug$ \cite{Giamarchi}.

\subsection{No Zeeman effect}

In the absence of magnetic field, one must distinguish between charge (c) and spin (s) excitations. The former propagate at the plasmon velocity $u=v_F/g$ whereas the latter rather propagate at the Fermi velocity $v_F$.

Fractional excitations from the charge sector are now explicitly described by the operators \cite{KV,KarynF}
\begin{equation}
{\cal C}_{\pm}^{Q_{c,\pm}}(x,t)=\exp\left[-i{\sqrt\frac{\pi}{2}}
Q_{c,\pm}\theta_{\pm}^c(x,t)\right],
\end{equation}
where $\theta_{\pm}^c = \theta_c\mp {\phi}_c/g$ similar to the spinless situation and the fractional charges are precisely given by $Q_{c,\pm} = (Q_{\uparrow}+Q_{\downarrow})/2 \pm g(J_{\uparrow}+J_{\downarrow})/2$. More precisely ($Q_{\alpha}$ and $J_{\alpha}$ are the charge and current for a given spin polarization),
\begin{eqnarray}
Q_{c+}+Q_{c-} &=& Q_{\uparrow}+Q_{\downarrow} \\ \nonumber
u(Q_{c+}-Q_{c-}) &=& v_F \left(J_{\uparrow}+J_{\downarrow}\right).
\end{eqnarray}
Moreover, spin excitations ${\cal S}_{\pm}$ are embodied by the operators \cite{KV,KarynF}:
\begin{equation}
{\cal S}_{\pm}^{Q_{s,\pm}}(x,t)=\exp\left[-i{\sqrt\frac{\pi}{2}}{Q_{s,\pm}}\theta_{\pm}^s(x,t)\right],
\end{equation}
where $\theta_{\pm}^s = \theta_s\mp {\phi}_s$ and $Q_{s,\pm} = (Q_{\uparrow}-Q_{\downarrow})/2 \pm (J_{\uparrow}-J_{\downarrow})/2$. Now, suppose we add in a spin-up electron 
going to the right. This is a mixed $Q_{\uparrow}=J_{\uparrow}=1$ and $Q_{\downarrow}=J_{\downarrow}=0$
excitation, which then will split into {\it three}
fractional parts. First, a right-moving 
charge $Q_{c,+}=(1+g)/2$ 
propagating at velocity $+u$ and a counter-propagating charge $Q_{c,-}=(1-g)/2$ 
with velocity $-u$. Second, a right-moving spinon with a spin component 
$S_+^z=Q_{s,+}/2=1/2$ propagating at $+v_F$.

This approach explains the Dzyaloshinskii-Larkin Green's function obtained via
Ward identities (at $T=0$) \cite{Larkin}
\begin{eqnarray}
\label{Larkin}
G_{+\uparrow}(x,t)\ = \frac{e^{i k_F x}}{2\pi}\frac{1}{x-v_F t+i0^+} \left(\frac{x-v_F t+i/q_o}{x-ut+i/q_o}\right)^{1/2}\left[{q_o}^2\left(x-ut+i/q_o\right)\left(x+ut-i/q_o\right)\right]^{-\frac{\gamma}{2}}.
\end{eqnarray}
Here, $q_o$ must be identified as a momentum cutoff.

\section{Calculation of $I_{\Phi}$}

Here, we provide a pedestrian derivation of the calculation of the flux-dependent part of the current in Eq. (45). In general, the tunneling Hamiltonian takes the form \cite{noteI}:
\begin{equation}
H_{tun}=\sum_{i=0,d}\Gamma_{i}u\Psi^{\dagger}_{2}(x=i)\Psi_{1}(x=i)+h.c.;
\end{equation}
Here, the lowerscripts $1,2$ refer to the upper and lower wire, respectively. In the relevant high-temperature regime $Td/u>1$ the most important contribution stems from the forward tunneling $\Gamma_i u\Psi_{2\pm}^{\dagger}(x=i)\Psi_{1\pm}(x=i)+h.c.$ as elucidated below.
%We will denote $\Gamma_{0}=\sqrt{T_0}\exp(i\mu_{1\pm}t/\hbar)$ and %$\Gamma_{d}=\sqrt{T_L}\exp(i\mu_{1\pm}t/\hbar)\exp(2i\pi\varphi)$ where $T_i$ denote the %transmission probabilities.  We will choose $\mu_{1+}=eV$ and $\mu_{1-}=0$;  another gauge would %essentially give the same physical result. 
Then, by exploiting our previous Ref. \cite{KLH2}, we converge to Eq. (44). Now, we can make use of
 \begin{equation}
<\Psi_{1+}(d,t)\Psi_{1+}^{\dagger}(0,0)>\ \approx -i\frac{a^{2\gamma}}{2\pi}
{\left[\frac{\pi}{u\beta}\right]}^{2\gamma+1}
\left[\sinh\left(\frac{\pi}{\beta}[t+\frac{d}{u}-i0^+]\right)\right]^{-\gamma}
\left[\sinh\left(\frac{\pi}{\beta}[t-\frac{d}{u}-i0^+]\right)\right]^{-\gamma-1},
\end{equation}
and similarly for $(-i)G_{2+}^<(-d,-t)=\ <\Psi_{2+}^{\dagger}(d,t)\Psi_{2+}(0,0)>$.
%\begin{equation}
%<T\Psi_{2+}^{\dagger}(d,t)\Psi_{2+}(0,0)>\ \approx i\frac{a^{2\gamma}}{2\pi}
%{\left[\frac{\pi}{u\beta}\right]}^{2\gamma+1}
%\left[\sinh\left(\frac{\pi}{\beta}[t+\frac{d}{u}-i0^+]\right)\right]^{-\gamma}
%\left[\sinh\left(\frac{\pi}{\beta}[t-\frac{d}{u}-i0^+]\right)\right]^{-\gamma-1}.
%\end{equation}
$X_{d0}^R(t)$ appearing in Eq. (44) can be computed in terms of products such as $G_{2+}^{<} G_{1+}^{>}$ along the lines of Ref. \onlinecite{Kindermann}. For free electrons, we can check that 
\begin{eqnarray}
I_{\Phi} = -2\frac{e^2}{h}|\Gamma_0\Gamma_d| \frac{1}{4} \Im m\int dt (-i)e^{iVt} e^{i2\pi\frac{\Phi}{\Phi_o}}P\left[\frac{1}{(t-d/u)^2}\right] +h.c. \approx  V\frac{e^2}{h}|\Gamma_0\Gamma_d| \cos(2\pi\Phi/\Phi_o).
\end{eqnarray}
Taking into account interactions between electrons, in the quantum realm $T\rightarrow 0$, results in
\begin{eqnarray}
I_{\Phi} \approx -2\frac{e^2}{h}|\Gamma_0\Gamma_d| \frac{a^{4\gamma}}{4u^{4\gamma}} \Im m\int dt (-i)e^{iVt} e^{i2\pi\frac{\Phi}{\Phi_o}}P\left[\frac{1}{t^{2+4\gamma}}\right] +h.c. \propto V\frac{e^2}{h}(|V|/D)^{4\gamma}|\Gamma_0\Gamma_d| \cos(2\pi\Phi/\Phi_o).
\end{eqnarray}
In the high-temperature limit $T\gg V$, one can make use of Eq. (42) leading to
\begin{equation}
<\Psi_{1+}(d,t)\Psi_{1+}^{\dagger}(0,0)>\ \approx \frac{1}{u} e^{ik_F d} e^{-d/u\tau_F}[T/D]^{2\gamma} \int _0^{+\infty} dE e^{iEd/u-iEt} = e^{ik_F d} [T/D]^{2\gamma} e^{-d/u\tau_F}\frac{i}{d-ut+i0^+}.
\end{equation}
It immediately follows that
\begin{eqnarray}
I_{\Phi} &=& -2\frac{e^2}{h}|\Gamma_0\Gamma_d| \frac{1}{4} [T/D]^{4\gamma}e^{-2d/u\tau_F} \Im m\int dt (-i)e^{iVt} e^{i2\pi\frac{\Phi}{\Phi_o}}P\left[\frac{1}{(t-d/u)^2}\right] +h.c. \\ \nonumber
&=&  V\frac{e^2}{h}|\Gamma_0\Gamma_d| [T/D]^{4\gamma} \cos(2\pi\Phi/\Phi_o)e^{-2d/u\tau_F}.
\end{eqnarray}
{\it The prominent attenuation of the interferences reflects the exponential decay of the Green
functions $G^>_+$ and $G^<_+$ in each wire.}
Finally, in the crossover region $V\sim T$, we rather estimate
\begin{eqnarray}
I_{\Phi} &\approx& -2\frac{e^2}{h}|\Gamma_0\Gamma_d| \frac{1}{4} D^{-4\gamma} T^{2\gamma}e^{-2d/u\tau_F}\Im m\int dt (-i)e^{iVt} e^{i2\pi\frac{\Phi}{\Phi_o}}P\left[\frac{1}{t^{2+2\gamma}}\right] +h.c. \\ \nonumber
&\approx& \frac{e^2}{h}V |\Gamma_0\Gamma_d| T^{2\gamma} |V|^{2\gamma} D^{-4\gamma}\cos(2\pi\Phi/\Phi_o)e^{-2d/u\tau_F}.
\end{eqnarray}
At this level, it is appropriate to compare the results with those of two counter-propagating edge states with $g=1$:
\begin{eqnarray}
I_{\Phi} &=& -2\frac{e^2}{h}|\Gamma_0\Gamma_d| \frac{1}{4} \Re e\int dt e^{iVt} e^{i2\pi\frac{\Phi}{\Phi_o}+i2k_F d}\left[\frac{1}{t-d/u-i0^+}\sinh^{-1}(\pi T 2d/u)\right]+h.c. 
\\ \nonumber
&\approx& \frac{e^2 V}{h} \frac{2\pi Td}{u}|\Gamma_0 \Gamma_d| \cos(2\pi\Phi/\Phi_o+2k_F d)\sinh^{-1}\left[\frac{\pi T 2d}{u}\right].
\end{eqnarray}
(In the probably unrealistic limit $Td/u\rightarrow 0$, the backscattering term gives the same current as the forward term.)

\end{appendix}


\begin{thebibliography}{10}
\bibitem{noteI}
   Compared to our Ref. \cite{KLH2}, we have redefined $\Gamma_{o,d}\rightarrow u\Gamma_{o,d}$ such that the tunneling amplitudes are dimensionless.
\bibitem{NoteP}
  This definition of $\Psi^{\dagger}_+$ is in agreement with the usual Wen's notatation: in Ref. \cite{Geller},  the momentum density canonically conjugate to $\phi_{\pm}$ is $\pm \partial_x \phi_{\pm}/(2\pi g)$ and thus one must carefully identify $\theta_{\pm}$ to $\phi_{\pm}/g$ and $Q_{\pm}=\pm 1$.
\end{thebibliography}
\end{document}